# On The Capacity Deficit of Mobile Wireless Ad Hoc Networks: A Rate Distortion Formulation

Nabhendra Bisnik, *Student Member, IEEE,* and Alhussein A. Abouzeid, *Member, IEEE*

## Abstract

Overheads incurred by routing protocols diminish the capacity available for relaying useful data in a mobile wireless ad hoc network. Discovering lower bounds on the amount of protocol overhead incurred for routing data packets is important for the development of efficient routing protocols, and for characterizing the actual (effective) capacity available for network users. This paper presents an information-theoretic framework for characterizing the minimum routing overheads of geographic routing in a network with mobile nodes. specifically, the minimum overhead problem is formulated as a rate-distortion problem. The formulation may be applied to networks with arbitrary traffic arrival and location service schemes. Lower bounds are derived for the minimum overheads incurred for maintaining the location of destination nodes and consistent neighborhood information in terms of node mobility and packet arrival process. This leads to a characterization of the deficit caused by the routing overheads on the overall transport capacity.

## Index Terms

Wireless Ad Hoc Networks, Geographic Routing, Routing Overheads, Capacity Deficit, Rate-Distortion Theory

## I. INTRODUCTION

**M**OBILE ad hoc networks are characterized by dynamically changing network topology which makes routing packets in an ad hoc network a very challenging problem. Routing protocols either fail to cope with the changing topology and yield low packet delivery rate or incur very high overhead. It is important to understand the minimum overhead incurred for routing packets with certain level of reliability. Knowledge of such a fundamental overhead limit would not only allow researchers to know how much a protocol deviates from the theoretical minimum, but also inspire the development of routing protocols that achieve the limit.

We view that the primary goal of a routing protocol is to gather and disseminate *state information* such that a node may take packet forwarding decisions that satisfy certain performance criteria. The state information may be comprised of link states, node locations, velocity and direction of nodes, queue lengths, etc. The performance criteria could be minimum delay, maximum throughput, maximum lifetime, delivering certain fraction of packets or simply best effort delivery of packets.

It is difficult to compare the overhead of various routing protocols since as yet there exist no absolute bounds on the minimum overhead incurred by the well known protocol classes. The premise of this work (as well as related prior work [1]) is that, instead of asking "what is the best routing protocol?" we could characterize the routing overheads incurred by a few key classes of protocols. The work in [1] addressed the overheads for proactive routing protocols. This paper considers geographic routing, which has become very commonly used as a network layer paradigm for multi-hop wireless networks.

In this paper we present an information-theoretic framework for characterizing the minimum overhead incurred by geographical routing protocols. Information theory provided us with lower bounds on the minimum number of bits required to encode a source. Thus, it is reasonable to expect that information theory would be a suitable tool for developing lower bounds on the amount of overhead incurred by routing protocols for disseminating and gathering state information in a mobile ad hoc network.

In geographic routing each node maintains its location information at one or more *location servers* (e.g. see [2], [3]). When a source wants to forward a packet to a destination, it queries an appropriate location server for the location of the destination. The location server replies to the source node with the available location information. Thereafter, the source and intermediate nodes forward the packet according to the location of the destination. It is pointed out in [4] that the fraction of packets delivered by geographical routing varies inversely with the average error in location information stored at the location servers. Thus, maintaining packet delivery ratio above a given threshold corresponds to maintaining location errors below a certain threshold.

We categorize geographic routing overheads into two categories: (i) *Location update overhead*: The overhead incurred in updating the location servers such that the location errors in the reply to location queries is less than $\epsilon$, and (ii) *Beacon overhead*: The overhead incurred in beacon transmission such that the probability that a node has consistent neighborhood information when it needs to forward a packet is greater than $1 - \delta$. We formulate the problems of finding the minimum values of the above-mentioned overheads as rate-distortion problems. For location update overheads, the distortion measure used is squared error in the location information stored at the location servers (*squared error distortion measure*). For beacon overheads, the distortion measure is the probability that a perceived neighbor is not an actual neighbor (*Hamming distortion measure*). Using a rate-distortion formulation, we present lower bounds on the minimum geographic routing overhead incurred in terms of node mobility, packet arrival process, and reliability criteria $\epsilon$ and $\delta$. First, we consider one-dimensional network case and then extend the results to two-dimensional networks.

We then unite the results obtained here on the minimum geographic routing overheads with the results on the transport capacity of stationary multihop wireless networks evaluated in [5], in order to characterize the effective capacity available for users. It is observed that when the node mobility is high and the average packet inter-arrival time is sufficiently small, the complete transport capacity of an ad hoc network may be consumed by routing overheads. We derive an upper bound on the critical network size above which all the transport capacity of the network would be consumed by the routing overheads and no useful communication would be possible. In this paper we only consider the scenario where the routing protocol initiates the forwarding process as soon as a packet is generated at the source. Thus the potential capacity improvement due to node mobility (achieved at the cost of delay associated with waiting for the destination to move to a nearby location) pointed out in [6], [7] is not applicable to our work.

### A. Main Contributions

The main contributions of this paper may be summarized as follows:

1) We present a new information-theoretic formulation for evaluating the minimum routing overhead incurred by geographic routing. The formulation is general so that it may be applied to any node distribution, packet arrival process, and may be extended to any location service scheme and mobility model.

2) For Brownian mobility model and various packet inter-arrival time distributions, we evaluate lower bounds on the minimum rate at which a node must transmit its location information and beacons such that the packets are routed with desired level of reliability. Combining both overheads, we find a lower bound on the capacity deficit caused by geographic routing overheads in mobile wireless ad hoc networks.

3) We characterize the effective transport capacity of an ad hoc network after taking into account the minimum routing overheads that must be incurred for reliable geographic routing.

4) For a given packet arrival process, standard deviation of Brownian motion and reliability parameters ($\epsilon$, $\delta$), we evaluate the upper bound on the number of nodes the ad hoc network can support such that the complete transport capacity of the network is not used up by routing overheads.

### B. Paper Outline

The rest of the paper is organized as follows. A brief overview of related work is presented in Section II. The network model is presented in Section III. The rate-distortion formulation and evaluation of a lower bound on the minimum position update and beacon overheads are presented in Sections IV and V respectively. A discussion of the capacity deficit caused by routing overheads is presented in Section VI. The application of the formulation to other scenarios and some possible extensions of the network model is discussed in Section VII. We present conclusions and directions for future research in Section VIII.

## II. RELATED WORK

So far, we believe information theory has not significantly influenced the design and understanding of communication network protocols (an opinion we share with the authors of [8]). One of the earliest (and most significant) attempts in using information theory to enhance the understanding of communication networks was made in [9]. Gallager [9] used an information-theoretic approach in order to characterize a lower bound on the amount of protocol information required to keep track of the sender, receiver and timing of messages for a simple (stationary) network model. It is found that although the introduction of message delay decreases the protocol information, small average message length and high message arrival rate may lead to prohibitively high protocol overhead.

A few relatively recent papers have used information theory to understand the effects of node mobility on wireless networks. An analytical framework, based on entropy of node location, for characterizing delay and overhead associated with paging and routing a call to a mobile station in a cellular environment is provided in [10]. The complexity of tracking a mobile user in a cellular environment is studied using a information-theoretic approach and a position update and paging scheme is proposed in [11]. An entropy based modeling framework for evaluating and supporting route stability in mobile ad hoc networks is proposed in [12]. In [13], the authors propose the entropy of link change as the metric for mobility models against which performance of wireless network protocols could be evaluated.

The overhead incurred by routing protocols and their scalability properties has been studied in [1], [14]–[23]. The initial studies [14], [15], [17], [18] are mainly simulation based. These studies point out that none of the routing protocols performs well across all scenarios. Instead each protocol performs well in some scenarios and bad in others.

Although simulation-based studies provide useful information about the performance of routing protocols, the observations may not be generalized to all scenarios. Therefore it is important to have analytical results for the performance of routing protocols in order to be able to develop a deep and general understanding of the trade-offs involved. In [16], [20], the authors present an analytical framework for characterizing the routing overhead for ad hoc routing protocols. Asymptotic results are provided for the overhead of proactive and reactive protocols in terms of network and routing protocol parameters such as packet arrival rate, hello packet transmission rate, hello packet size, size of route request packet, topology broadcast rate etc. The overheads of specific routing protocols were modeled in [19]. The impact of the routing-layer traffic patterns (in terms of number of hops between source-destination pairs) on the scalability of reactive routing protocols in ad hoc networks with unreliable but stationary nodes is studied in [22], [23]. It is found that the reactive routing protocols may scale infinitely (i.e. the routing overhead does not tend to infinity) with respect to network size for some traffic patterns. The analysis is extended to and similar results are obtained for cluster-based routing algorithms in [21].

Our work is along the lines of [1] where the authors use an information-theoretic approach to characterize the minimum routing overhead and memory requirements of topology-based (proactive) hierarchical routing protocols for ad hoc networks. The entropy of ad hoc network topologies as well as the entropy rate are used in [1] to find the above mentioned bounds. However, here, the family of routing protocols considered is geographic routing protocols and the performance constraints are also taken into account. This leads to a new problem formulation as rate distortion which was not considered in earlier work and new results on the effect on transport capacity. Some preliminary results of our work appeared in [24]. Several significant contributions are made by this paper in addition to [24]. In this paper we analyze beacon overhead for various packet arrival processes, evaluate tighter lower bounds on location update overhead, extend the overhead analysis to other scenarios of geographic routing, and include new results of capacity deficit obtained through numerical analysis.

## III. NETWORK MODEL

The network consists of $n$ mobile nodes. The nodes perform Brownian motion with variance $\sigma^2$. We consider two kinds of network deployments: (i) *One dimensional case:* nodes located along a circle of perimeter $L$, and (ii) *Two dimensional case:* nodes located over a torus of surface area $A$. The central and lateral radii of the torus are denoted by $R_c$ and $R_l$ respectively.

The closed curve and surface are chosen for the study, instead of a finite line or a rectangle, in order to avoid the complexity of modeling the behavior of Brownian motion at boundary points.

We assume that $L >> \sigma^2$ and $R_c, R_l >> \sigma^2$. The large dimensions ensure that the nodes do not wrap around the curve or surface during small intervals of time. So if we look at the motion of a node during a small interval of time, then with probability almost one the motion is similar to Brownian motion on an infinite line or plane with the initial node position as the origin. Thus, in the rest of the paper, we treat the motion of nodes during the time scale corresponding to packet inter-arrival times as motion on a plane or straight line. Over time the nodes do not drift apart from each other, as they would on a infinite line or plane, but just keep moving around on the circle or the torus.

Conversely, this may be viewed as if we are observing the Brownian motion of the nodes on an infinite line or plane and mapping their positions back on the circle or torus, respectively. For example, consider a node that performs Brownian motion along the x-axis and whose initial position is the origin. At time $t$, suppose the node is located at $X(t)$. Then it may be mapped to a point $\mod_L(X(t))$ away from the initial position of the node on the circle, with distance measured in counter-clockwise direction. Similar mapping is possible in the case of torus by considering an infinite plane. Thus instead of keeping track of the positions of nodes on the circle or torus, we use the coordinates of the nodes on x-axis and infinite plane. This scheme works since we are only interested in the change in positions of nodes during packet inter-arrival periods. The coordinates of nodes are denoted by $X_i(t)$. Hence $X_i(t) = \{X_{i1}(t)\}$ and $X_i(t) = \{X_{i1}(t), X_{i2}(t)\}$ for one and two-dimension case respectively. The location information of node $i$ available at the location server at time $t$ is denoted by $\hat{X}_i(t)$, hence, $\hat{X}_i(t) = \{\hat{X}_{i1}(t)\}$ and $\hat{X}_i(t) = \{\hat{X}_{i1}(t), \hat{X}_{i2}(t)\}$ for one and two-dimension case respectively.

The $j^{th}$ packet destined to destination $i$ generated at a node (source of the $j^{th}$ packet) in the network at time $T_i(j)$, $\forall$ $j \geq 1$. Define $T_i(0) \triangleq 0 \ \forall \ 1 \leq i \leq n$. For all $j \geq 1$, let $S_j \triangleq T_i(j) - T_i(j-1)$ denote the packet inter-arrival time which is independently and identically distributed (i.i.d.) with pdf $f_S(t)$, such that $f_S(t) = 0 \ \forall t < 0$ and $E[S]$ exists. Similarly let $\tau_i(k)$ denote the time at which the $k^{th}$ packet is forwarded by node $i$, with $\tau_i(0) \triangleq 0$. The forwarded packets include both the packets generated by node $i$ and the packets for which the node acts as an intermediate relaying node. The inter-arrival time of the forwarded packets, $\tau_i(k+1) - \tau_i(k) \ \forall \ k > 0$, is an i.i.d random variable whose pdf denoted by $f_\tau(t)$.

We assume a GHLS [25] like location service scheme. For each node, a hashing function is used to map the node id to a *home region* within the deployment area. A node within the home region acts as the location server for the node. A node sends updates its location information by sending update packets destined to its home region. The packets are forwarded using geographic routing. The location server, located within the home region, receives these update packets and maintains the location information of the node. Since the nodes are mobile, the members of home region change over time. A lightweight hand-off mechanism, proposed in [25], may be used to hand-over the task of maintaining location information to new nodes when a location server leaves the home region. The hand-off mechanism does not require transmission of extra control messages. The current location server uses only the beacons received from neighbors in order to ascertain if it needs to hand-over the responsibility of maintaining the location information. We discuss extensions of our work to other location service scheme models in Section VII.

The communication radius of each node is $r$ meters. When a new packet generated at a source node, it queries the location server of the packet destination for the location of the destination. The packet is routed to the destination according to greedy geographic forwarding using the destination location information returned by the location server. It is assumed the position of a destination does not change significantly while the location server is being queried by the source and the packet is being forwarded through the network. In other words the time scale of forwarding a packet is much smaller than that required for a significant change in position. Also the network is assumed to be always connected such that nodes can communicate with the desired location servers.

## IV. LOCATION UPDATE OVERHEAD

In this section we evaluate a lower bound on the minimum rate at which a node must transmit its location information such that the average error in its location stored at the location server is less than $\epsilon$ whenever the server is queried. We first introduce the notation and rate-distortion formulation, followed by analysis for one-dimensional and two-dimensional networks. We also evaluate lower bounds for deterministic, uniform and exponential packet arrival processes.

### A. Notation and Rate-Distortion Formulation

*Definition 1:* $D_i(t)$ is the squared-error in the location information of destination $i$ available at its location server at time $t$, i.e.,

$$D_i(t) = |X_i(t) - \hat{X}_i(t)|^2 \tag{1}$$

where $|X_i(t) - \hat{X}_i(t)| = \sqrt{\sum_{j=1}^{m} \left( X_{ij}(t) - \hat{X}_{ij}(t) \right)^2}$.

*Definition 2:* $X_i^N = \{X_i(T_1), X_i(T_1), \ldots, X_i(T_N)\}$ is the vector of locations of destination $i$ at time instances $T_j$, $1 \le j \le N$. Similarly $\hat{X}_i^N = \{\hat{X}_i(T_1), \hat{X}_i(T_2), \ldots, \hat{X}_i(T_N)\}$ is the vector of location information at the location server of destination $i$ at time instances $T_j$, $1 \le j \le N$.

*Definition 3:* $\mathcal{X}_i^N$ and $\hat{\mathcal{X}}_i^N$ are sets of all possible vectors $X_i^N$ and $\hat{X}_i^N$, respectively.

*Definition 4:* $P_N[x_i^N; \hat{x}_i^N]$ denotes the probability that $X_i^N = x_i^N$ and $\hat{X}_i^N = \hat{x}_i^N$, where $x_i^N \in \hat{\mathcal{X}}_i^N$ and $\hat{x}_i^N \in \mathcal{X}_i^N$.

*Definition 5:* $\overline{D}_{iN}$ is defined as

$$\overline{D}_{iN} \triangleq \frac{1}{N} \sum_{j=1}^{N} E[D_i(T_j)] \tag{2}$$

where $E[D_i(T_j)]$ is given by

$$E[D_i(T_j)] = \sum_{x_i^N \in \mathcal{X}_i^N} \sum_{\hat{x}_i^N \in \hat{\mathcal{X}}_i^N} P_N[x_i^N; \hat{x}_i^N] D_i(T_j) \tag{3}$$

*Definition 6:* $\mathcal{P}_N(\epsilon^2)$ is defined as the family of probability distribution functions $P_N[x_i^N; \hat{x}_i^N]$ for which $\overline{D}_{iN} \le \epsilon^2$.

*Definition 7:* $R_N(\epsilon^2)$ is defined as the $N^{\text{th}}$-order rate-distortion function – the minimum rate at which a destination must transmit the location information such that the $\overline{D}_{iN} \le \epsilon$. According to [26], $R_N(\epsilon^2)$ is given by

$$R_N(\epsilon^2) = \min_{P_N \in \mathcal{P}_N(\epsilon^2)} \frac{1}{N} I_{P_N}(X_i^N; \hat{X}_i^N) \tag{4}$$

where $I_{P_N}(X_i^N; \hat{X}_i^N)$ is the mutual information between $X_i^N$ and $\hat{X}_i^N$.

The minimum rate at which a destination must update its location information such that a large fraction of packets are delivered, represented by $R(\epsilon^2)$, is given by

$$R(\epsilon^2) = \lim_{N \to \infty} \min R_N(\epsilon^2) \tag{5}$$

### B. Location Overhead for One-Dimensional Networks

*Lemma 1:* The mutual information between $X_i^N$ and $\hat{X}_i^N$ satisfies the following relationship

$$\inf_{P_N \in \mathcal{P}_N} I_{P_N}(X_i^N; \hat{X}_i^N) \ge N R_1(\epsilon^2) \tag{6}$$

*Proof:* This proof is very similar to the proof of [9, Theorem 3].

Using the standard definition of mutual information we get

$$I_{P_N}(X_i^N; \hat{X}_i^N) = H(X_i^N) - H(X_i^N | \hat{X}_i^N) \tag{7}$$

Now consider $H(X_i^N | \hat{X}_i^N)$

$$H(X_i^N | \hat{X}_i^N) = H(X_{i1}(T_1) | \hat{X}_i^N) - \sum_{j=2}^{N} H(X_{i1}(T_j) | X_{i1}(T_1), X_{i1}(T_1), \ldots, X_{i1}(T_{j-1}), \hat{X}_i^N)$$

Since conditioning cannot increase the entropy, we have

$$H(X_i^N | \hat{X}_i^N) \le H(X_{i1}(T_1) | \hat{X}_i^N) + \sum_{j=2}^{N} H(X_{i1}(T_j) | X_{i1}(T_{j-1}) \hat{X}_{i1}(T_j)) \tag{8}$$

$$= H(X_{i1}(T_1) | \hat{X}_i^N) + \sum_{j=2}^{N} H(X_{i1}(T_j) - X_{i1}(T_{j-1}) | X_{i1}(T_{j-1}), \hat{X}_{i1}(T_j)) \tag{9}$$

(9) follows from (8) since $H(X_{i1}(T_j)|X_{i1}(T_{j-1})\hat{X}_{i1}(T_j))$ is already conditioned on $X_{i1}(T_{j-1})$, subtracting it from $X_{i1}(T_j)$ is similar to translating the random variable by a scalar. Define a new random variable, $Y_j$ $(1 \le j \le N-1)$, such that

$$Y_j = \hat{X}_{i1}(T_j) - X_{i1}(T_{j-1}) \tag{10}$$

Since $Y_j$ depends only on $\hat{X}_{i1}(T_j)$ and $X_{i1}(T_{j-1})$, we have

$$
\begin{aligned}
H(X_{i1}(T_j) - X_{i1}(T_{j-1})|X_{i1}(T_{j-1}), \hat{X}_{i1}(T_j)) &= H(X_{i1}(T_j) - X_{i1}(T_{j-1})|Y_j, X_{i1}(T_{j-1}), \hat{X}_{i1}(T_j)) \tag{11} \\
&\le H(X_{i1}(T_j) - X_{i1}(T_{j-1})|Y_j) \tag{12}
\end{aligned}
$$

Equation (12) follows from (11) because conditioning does not increase entropy. Thus we get the following upper bound on $H(X_i^N|\hat{X}_i^N)$

$$H(X_i^N|\hat{X}_i^N) \le H(X_{i1}(T_1)|\hat{X}_{i1}(T_1)) + \sum_{j=2}^{N} H(X_{i1}(T_j) - X_{i1}(T_{j-1})|Y_j). \tag{13}$$

Now consider $H(X_i^N)$, since $X_{i1}(T_j) - X_{i1}(T_{j-1})$ are independent of each other, we have

$$H(X_i^N) = H(X_{i1}(T_1)) + \sum_{j=2}^{N} H(X_{i1}(T_j) - X_{i1}(T_{j-1})) \tag{14}$$

Combining (7), (13) and (14), we get

$$I_{P_N}(X_i^N; \hat{X}_i^N) = I(X_{i1}(T_1); \hat{X}_{i1}(T_1)) + \sum_{j=2}^{N} I(X_{i1}(T_j) - X_{i1}(T_{j-1}); Y_j) \tag{15}$$

Notice that, the squared error in the location information may also be written as

$$D_i(T_j) = |Y_j - (X_{i1}(T_j) - X_{i1}(T_{j-1}))|^2 = |X_i(T_j) - \hat{X}_i(T_j)|^2 \tag{16}$$

Let $d_j = E[D_i(T_j)]$. Since $X_{i1}(T_1)$ has the same distribution as $X_{i1}(T_j) - X_{i1}(T_{j-1})$, and (16) is satisfied, therefore we have

$$I(I(X_{i1}(T_j) - X_{i1}(T_{j-1}); Y_j) \ge R_1(d_j) \; \forall \; j \ge 2 \tag{17}$$

Define $d_i \triangleq E[|X_{i1}(T_1) - \hat{X}_{i1}(T_1)|^2]$, then by substituting (17) in (15) and using the convexity of the rate distortion function $R_1$, we get

$$I_{P_N}(X_i^N; \hat{X}_i^N) \ge \sum_{j=1}^{N} R_1(d_j) \ge N R_1 \left( \frac{1}{N} \sum_{j=1}^{N} d_j \right) \tag{18}$$

Now since $P_N \in \mathcal{P}_N$, $(1/N) \sum_{j=1}^{N} d_j = \overline{D}_{iN} \le \epsilon^2$, therefore from (18) we have $I_{P_N}(X_i^N; \hat{X}_i^N) \ge N R_1(\epsilon^2)$. ∎

¿From Lemma 1 and the definition of rate distortion function (4) and (5) it follows that

$$R(\epsilon^2) \ge R_1(\epsilon^2) \tag{19}$$

The following theorem provides a lower bound on the minimum rate at which a destination must update its location information.

*Theorem 1:* The lower bound on the location update rate (in bits per packet) is given by

$$R(\epsilon^2) \ge h(X_{i1}(T_1)) - \frac{1}{2} \log 2\pi e \epsilon^2 \tag{20}$$

where $h(X_{i1}(T_1))$ is the differential entropy of the location of destination $i$ at the time when the first packet destined to it is generated in the network.

*Proof:* ¿From (19) we know that the minimum update rate is bounded by $R_1(\epsilon^2)$, which in turn is defined as

$$R_1(\epsilon^2) = \inf_{P_1 \in \mathcal{P}_1} I_{P_1}(X_{i1}(T_1); \hat{X}_{i1}(T_1))$$

Now consider $I_{P_1}(X_{i1}(T_1); \hat{X}_{i1}(T_1))$,

$$
\begin{aligned}
I_{P_1}(X_{i1}(T_1); \hat{X}_{i1}(T_1)) &= h(X_{i1}(T_1)) - h(X_{i1}(T_1) | \hat{X}_{i1}(T_1)) \\
&= h(X_{i1}(T_1)) - h(X_{i1}(T_1) - \hat{X}_{i1}(T_1) | \hat{X}_{i1}(T_1)) \qquad (21) \\
&\geq h(X_{i1}(T_1)) - h(X_{i1}(T_1) - \hat{X}_{i1}(T_1)) \qquad (22) \\
&\geq h(X_{i1}(T_1)) - h(X_{i1}(\mathcal{N}(0, E[(X_{i1}(T_1) - \hat{X}_{i1}(T_1))^2])) \qquad (23) \\
&\geq h(X_{i1}(T_1)) - \frac{1}{2} \log\left(2\pi e \epsilon^2\right) \qquad (24)
\end{aligned}
$$

Here (22) follows from (21) since conditioning does not increase entropy, (23) follows from (22) since for a fixed variance, the normal distribution has the highest differential entropy, and (24) follows from (23) since for $P_1 \in \mathcal{P}_1$, $E[(X_{i1}(T_1) - \hat{X}_{i1}(T_1))^2] \leq \epsilon^2$. Thus,

$$
R_1(\epsilon^2) \geq h(X_{i1}(T_1)) - \frac{1}{2} \log\left(2\pi e \epsilon^2\right) \qquad (25)
$$

and (20) follows directly from it. ∎

It should be noted that in under some situation $h(X_{i1}(T_1))$ may be less than $\frac{1}{2} \log\left(2\pi e \epsilon^2\right)$. Such situations may occur when $\sigma^2$ and/or $E[S]$ is small. Under these circumstances, the change in position of node between two packet generation instances may be comparable to the fidelity criterion ($\epsilon^2$) and hence small number of bits, if any, may be required to represent the change in position of a node between packet generation instances. However, over the course of time the position of a node may change appreciably which may require it to update its location information. When $h(X_{i1}(T_1)) < \frac{1}{2} \log\left(2\pi e \epsilon^2\right)$, the right hand side of of inequality (20) is meaningless. A more appropriate inequality will be

$$
R(\epsilon^2) \geq \max\left(h(X_{i1}(T_1)) - \frac{1}{2} \log\left(2\pi e \epsilon^2\right), 0\right) \qquad (26)
$$

We discuss the case where $h(X_{i1}(T_1)) < \frac{1}{2} \log\left(2\pi e \epsilon^2\right)$ in detail in Section VII.

Theorem 1 implies that the minimum update rate largely depends on $h(X_{i1}(T_1))$, which in turn depends on two factors: (i) the mobility pattern of the destination node and (ii) the packet inter-arrival process. Let $f_{X_1}(x)$ denote the pdf of $X_{i1}(T_1)$ (without loss of generality, $X_{i1}(T_0) = 0$). For Brownian motion with variance $\sigma^2$ and packet inter-arrival time distribution $f_S(t)$, $f_{X_1}(x)$ is given by

$$
f_{X_1}(x) = \int_{\tau=0}^{\infty} \frac{1}{\sqrt{2\pi\sigma^2\tau}} e^{-\frac{x^2}{2\sigma^2\tau}} f_S(\tau) d\tau \qquad (27)
$$

and $h(X_{i1}(T_1))$ is given by

$$
h(X_{i1}(T_1)) = -\int_{x=-\infty}^{\infty} f_{X_1}(x) \log\left(f_{X_1}(x)\right) dx \qquad (28)
$$

The lower bound on the minimum overhead incurred by location update information in bits/second, denoted by $U(\epsilon^2)$, is given by

$$
U(\epsilon^2) \geq \frac{1}{E[S]} \max\left(h(X_{i1}(T_1)) - \frac{1}{2} \log\left(2\pi e \epsilon^2\right), 0\right) \qquad (29)
$$

In the remaining part of this subsection we evaluate $f_{X_1}(x)$ and $R(\epsilon^2)$ for deterministic, uniform and exponential inter-arrival time distributions.

*1) Deterministic inter-arrival distribution:* We first consider the case of deterministic arrival case because it is easy to obtain closed form results for this case and therefore we can focus on gaining insights without bothering much about complicated analysis.

Suppose that packet destined to destination $i$ is generated in the network after every $T$ seconds, that is

$$
f_S(t) = \delta(t - T)
$$

For such an arrival process, $f_{X_1}(x)$ is given by

$$
f_{X_1}(x) = \frac{1}{\sqrt{2\pi\sigma^2 T}} e^{\frac{-x^2}{2\sigma^2 T}}
$$

and $h(X_{i1}(T_1))$ is given by

$$h(X_{i1}(T_1)) = \frac{1}{2}\log\left(2\pi e\sigma^2 T\right)$$

Therefore the minimum update rate in bits per packet for deterministic packet inter-arrival time is given by

$$R(\epsilon^2) \geq \max\left(\frac{1}{2}\log\left(\frac{\sigma^2 T}{\epsilon^2}\right), 0\right) \text{ bits/packet} \tag{30}$$

For the deterministic inter-arrival time the lower bound is similar to the minimum number of bits required to represent a Gaussian random variable with variance $\sigma^2 T$ subject to the constraint that the expected squared-error is less than $\epsilon^2$. This is because the change in position of a node during a packet inter-arrival interval is indeed a Gaussian random variable with variance $\sigma^2 T$.

The equation (30) bounds the minimum number of bits that a destination node must send to the location server for each packet destined to it that is generated in the network. For the deterministic arrival, the the packet arrival rate equals $1/T$ packets per second, thus minimum update rate in bits/second, represented by $U(\epsilon^2)$, is given by

$$U(\epsilon^2) \geq \frac{1}{2T}\max\left(\log\left(\frac{\sigma^2 T}{\epsilon^2}\right), 0\right) \text{ bits/second} \tag{31}$$

Notice that $R(\epsilon^2)$ increases with both $\sigma^2$ and $T$. This is because larger $\sigma^2$ and $T$ would imply larger uncertainty in the change in position of the destination during the packet arrival duration and hence more bits are required to represent the destination's position for each new packet. $U(\epsilon^2)$ also increases with $\sigma^2$ for the same reason. However, $U(\epsilon^2)$ decreases with increase in $T$. This is because the increase in update rate due to higher uncertainty associated with high $T$ is over-compensated by the fact that larger $T$ implies that updates have to be made less often.

For deterministic inter-arrival time it is also quite easy to construct a strategy to update the location server that achieves the bound in (31). The following strategy achieves the bound: Each destination updates its location at time $kT$ $(k = 0, 1, \ldots)$ seconds by encoding $\mathcal{N}(0, \sigma^2 T)$ Gaussian random variable corresponding to the change of position since last update such that the squared error distortion is less than $\epsilon^2$. However in a real-life scenario the arrival times of packets destined to a particular destination is not known a priori, although we might have some estimate of the arrival rate.

*2) Uniform inter-arrival time distribution:* Among all continuous time distributions with a given finite base and mean the uniform distribution has the highest entropy. Thus uniform distribution maximizes the uncertainty of packet arrival instances and thus would lead to maximum position update rate among all distributions with the same finite base and mean.

Consider inter-arrival time to be uniformly distributed between $[0, T]$ such that the inter-arrival time distribution is given by

$$f_S(t) = \begin{cases} \frac{1}{T}, & 0 \leq t \leq T \\ 0, & \text{otherwise} \end{cases} \tag{32}$$

From (27), $f_{X_1}(x)$ is given by

$$\begin{aligned} f_{X_1}(x) &= \frac{1}{T}\int_{\tau=0}^{T}\frac{1}{\sqrt{2\pi\sigma^2\tau}}e^{-\frac{x^2}{2\sigma^2\tau}}d\tau \\ &= \sqrt{\frac{2}{\pi\sigma^2 T}}e^{-\frac{x^2}{2\sigma^2 T}} + \frac{x}{\sigma^2 T}\text{erf}\left(\frac{x}{\sqrt{2\sigma^2 T}}\right) - \frac{|x|}{\sigma^2 T} \end{aligned} \tag{33}$$

Let $h_U$ denote the differential entropy of $f_{X_1}(x)$, then the update rate per packet is given by

$$R(\epsilon^2) \geq \max\left(h_U - \frac{1}{2}\log\left(2\pi e\epsilon^2\right), 0\right) \text{ bits/packet} \tag{34}$$

and the update rate in bits per second is given by

$$U(\epsilon^2) \geq \frac{2}{T}\max\left(h_U - \frac{1}{2}\log\left(2\pi e\epsilon^2\right), 0\right) \text{ bits/second} \tag{35}$$

*3) Exponential inter-arrival time distribution:* The motivation for considering the exponential distribution is that, among all the continuous time distribution with base $[0, \infty)$ and a given mean, the exponential distribution has the highest entropy. Also, the exponential distribution is widely used to model external inter-arrival time of packets to a given destination.

We consider an exponential distribution with mean $1/\alpha$, given by

$$f_S(t) = \alpha e^{-\alpha t}$$

¿From to (27), $f_{X_1}(x)$ is then given by

$$
\begin{aligned}
f_{X_1}(x) &= \int_0^\infty \frac{1}{\sqrt{2\pi\sigma^2\tau}} e^{-\frac{x^2}{2\sigma^2\tau}} \alpha e^{-\alpha\tau} d\tau \\
&= \int_0^\infty \frac{\alpha}{\sqrt{2\pi\sigma^2\tau}} e^{-\frac{x^2}{2\sigma^2\tau} - \alpha\tau} d\tau
\end{aligned}
\tag{36}
$$

Unfortunately a closed-form expression for the above integral cannot be evaluated. So we will use numerical methods to calculate its value.

Let $h_E$ be the differential entropy of the distribution $f_{X_1}(x)$ given in (36). Then the lower bounds on update rate per packet and update rate per second are given by

$$R(\epsilon^2) \geq \max\left(h_E - \frac{1}{2}\log\left(2\pi e\epsilon^2\right), 0\right) \text{ bits/packet} \tag{37}$$

$$U(\epsilon^2) \geq \alpha \max\left(h_E - \frac{1}{2}\log\left(2\pi e\epsilon^2\right), 0\right) \text{ bits/second} \tag{38}$$

*C. Location Overhead for Two-Dimensional Networks*

In this section we present the update rate analysis for two-dimensional networks, which is based on the analysis for one-dimensional case. We also evaluate the lower bound for various packet arrival processes and discuss the effect of arrival processes on the minimum update rate.

Brownian motion in two-dimensional space may be decomposed into two independent one dimensional Brownian motions along $x$ and $y$ coordinates each with a variance $\sigma^2/2$. Thus if $X_i(t) = \{X_{i1}(t), X_{i2}(t)\}$ denote the coordinates of destination $i$ at time $t$ then the distribution of $X_{i1}(t)$ is independent of the distribution of $X_{i2}(t)$. The following Lemma expresses $R\left(\epsilon^2\right)$ in terms of components corresponding to the two coordinates.

*Lemma 2:* For two-dimensional networks, the rate distortion $R\left(\epsilon^2\right)$ function may be written as

$$R(\epsilon^2) = \min_{0 \leq k \leq \epsilon} R^{(1)}(k^2) + R^{(2)}(\epsilon^2 - k^2) \tag{39}$$

where

$$R^{(1)}(\epsilon^2) = \lim_{N \to \infty} \inf_{P_N \in \mathcal{P}_N(k^2)} \frac{1}{N} I_{P_N}(X_{i1}^N; \hat{X}_{i1}^N) \tag{40}$$

$$R^{(2)}(\epsilon^2) = \lim_{N \to \infty} \inf_{P_N \in \mathcal{P}_N(\epsilon^2 - k^2)} \frac{1}{N} I_{P_N}(X_{i2}^N; \hat{X}_{i2}^N) \tag{41}$$

*Proof:* Recall the rate distortion function is

$$R(\epsilon^2) = \lim_{N \to \infty} \inf_{P_N \in \mathcal{P}_N(\epsilon^2)} \frac{1}{N} I_{P_N}(X_i^N; \hat{X}_i^N)$$

Now consider $I_{P_N}(X_i^N; \hat{X}_i^N)$. For two dimensional networks, this may be written as

$$
\begin{aligned}
I_{P_N}(X_i^N; \hat{X}_i^N) &= I_{P_N}(X_{i1}^N, X_{i2}^N; \hat{X}_i^N) \\
&= I_{P_N}(X_{i1}^N; \hat{X}_i^N) + I_{P_N}(X_{i2}^N; \hat{X}_i^N | X_{i1}^N) \\
&= I_{P_N}(X_{i1}^N; \hat{X}_i^N) + I_{P_N}(X_{i2}^N; \hat{X}_i^N) \\
&= I_{P_N}(X_{i1}^N; \hat{X}_{i1}^N) + I_{P_N}(X_{i1}^N; \hat{X}_{i2}^N | \hat{X}_{i1}^N) + I_{P_N}(X_{i2}^N; \hat{X}_{i2}^N) + I_{P_N}(X_{i2}^N; \hat{X}_{i1}^N | \hat{X}_{i2}^N) \\
&= I_{P_N}(X_{i1}^N; \hat{X}_{i1}^N) + I_{P_N}(X_{i2}^N; \hat{X}_{i2}^N)
\end{aligned}
\tag{42}
$$

where $X_{i1}^N = \{X_{i1}(T_1), X_{i1}(T_2), \ldots, X_{i1}(T_N)\}$, $X_{i2}^N = \{X_{i2}(T_1), X_{i2}(T_2), \ldots, X_{i2}(T_N)\}$, $\hat{X}_{i1}^N = \{\hat{X}_{i1}(T_1), \hat{X}_{i1}(T_2), \ldots, \hat{X}_{i1}(T_N)\}$ and $\hat{X}_{i2}^N = \{\hat{X}_{i2}(T_1), \hat{X}_{i2}(T_2), \ldots, \hat{X}_{i2}(T_N)\}$.

We know that $\overline{D}_{iN} \le \epsilon^2$ implies

$$
\frac{1}{N}\sum_{j=1}^{N} E\left[(X_{i1}(T_j) - \hat{X}_{i1}(T_j))^2\right] + \frac{1}{N}\sum_{j=1}^{N} E\left[(X_{i2}(T_j) - \hat{X}_{i2}(T_j))^2\right] \le \epsilon^2
$$

The distortion constraint is satisfied if $\frac{1}{N}\sum_{j=1}^{N} E\left[(X_{i1}(T_j) - \hat{X}_{i1}(T_j))^2\right] \le k^2$ and $\frac{1}{N}\sum_{j=1}^{N} E\left[(X_{i2}(T_j) - \hat{X}_{i2}(T_j))^2\right] \le \epsilon^2 - k^2$. Combining this and (42), we get

$$
R(\epsilon^2) = \min_{0 \le k \le \epsilon} \lim_{N \to \infty} \inf_{P_N \in \mathcal{P}_N(k^2)} \frac{1}{N} I_{P_N}(X_{i1}^N; \hat{X}_{i1}^N) + \inf_{P_N \in \mathcal{P}_N(\epsilon^2 - k^2)} \frac{1}{N} I_{P_N}(X_{i2}^N; \hat{X}_{i2}^N)
\tag{43}
$$

which leads to (39).                                                                                                        ∎

¿From Lemma 1 and Theorem 1, it follows that

$$
R^{(1)}(k^2) \ge h(X_{i1}(T_1)) - \log\left(2\pi e k^2\right)
\tag{44}
$$

$$
R^{(2)}(\epsilon^2 - k^2) \ge h(X_{i2}(T_1)) - \log\left(2\pi e(\epsilon^2 - k^2)\right)
\tag{45}
$$

¿From (44) and (45), it is clear that the right hand side of (39) is minimized for $k^2 = \epsilon^2/2$. This leads to the following theorem.

*Theorem 2:* In order to ensure that the average error in location information used for forwarding packets is less than $\epsilon$, the lower bound on the location update rate (in bits per packet) for a two-dimensional network is given by

$$
R(\epsilon^2) \ge \max\left(h(X_{i1}(T_1)) + h(X_{i2}(T_1)) - \log\left(\pi e \epsilon^2\right), 0\right) \text{ bits/packet}
\tag{46}
$$

and the overhead incurred in bits/sec ($U(\epsilon)$) is given by

$$
U(\epsilon^2) \ge \frac{1}{E[S]} \max\left(h(X_{i1}(T_1)) + h(X_{i2}(T_1)) - \log\left(\pi e \epsilon^2\right), 0\right) \text{ bits/sec}
\tag{47}
$$

We now derive the lower bounds for deterministic, uniformly distributed and exponentially distributed inter-arrival times.

*1) Deterministic packet arrival:* For the deterministic packet arrival process, where packets arrive at $t = kT$, $k = 1, 2, \ldots \infty$, the probability distribution functions of $X_{i1}(T_1)$ and $X_{i2}(T_1)$ are given by

$$
f_{X_1}(x) = f_{X_2}(x) = \frac{1}{\sqrt{\pi \sigma^2 T}} e^{-\frac{x^2}{\sigma^2 T}}
\tag{48}
$$

and $h(X_{i1}(T_1))$ and $h(X_{i2}(T_1))$ is given by

$$
h(X_{i1}(T_1)) = h(X_{i2}(T_1)) = \frac{1}{2} \log\left(\pi e \sigma^2 T\right)
$$

Thus the lower bound on location update rate in bits/packet and bits/second is given by

$$
R(\epsilon^2) \ge \max\left(\log\left(\frac{\sigma^2 T}{\epsilon^2}\right), 0\right) \text{ bits/packet}
\tag{49}
$$

$$
U(\epsilon^2) \ge \frac{1}{T}\left(\log\left(\frac{\sigma^2 T}{\epsilon^2}\right), 0\right) \text{ bits/second}
\tag{50}
$$

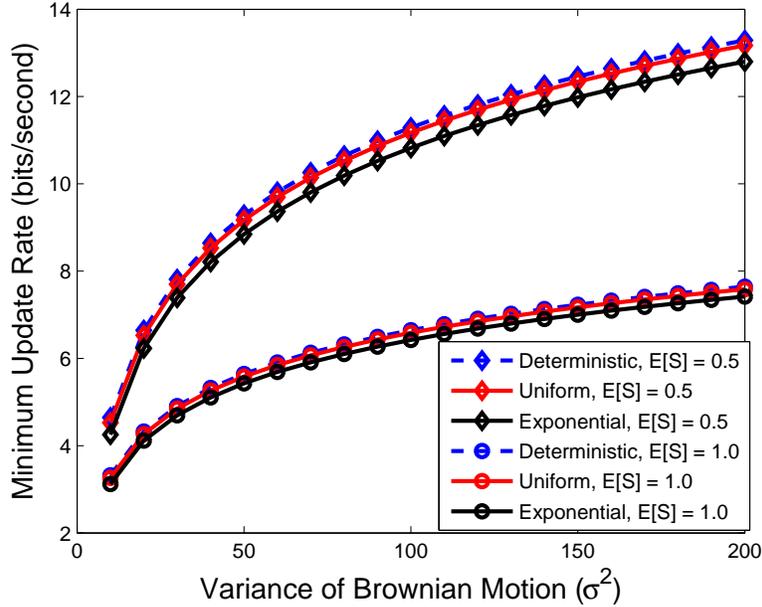

Fig. 1.   Update rate vs variance of Brownian motion.

The behavior of $R(\epsilon^2)$ and $U(\epsilon^2)$ is similar to that observed in the one-dimensional case. In fact the minimum rate is simply the double of what was observed in the one dimensional case. This is not surprising since the change in x-coordinate of the destination node is independent of the change in the y-coordinate of the destination. The change has variance $\sigma^2/2$ rather than $\sigma^2$ as in the case of one-dimensional motion thus one might expect the minimum rate in 2-D to be less than 2 times the minimum rate in 1-D case. However this decrease in entropy due to decreased variance is compensated by the fact that allowable squared error in both the coordinates is also decreased, leading to a factor of 2.

*2) Uniform distribution of packet inter-arrival time:*  For the uniform packet arrival process described in (32), the probability distribution functions of $X_{i1}(T_1)$ and $X_{i2}(T_1)$ are given by

$$f_{X_1}(x) = f_{X_2}(x) = \sqrt{\frac{4}{\pi\sigma^2 T}}e^{-\frac{x^2}{\sigma^2 T}} + \frac{2x}{\sigma^2 T}\mathrm{erf}\left(\frac{x}{\sqrt{\sigma^2 T}}\right) - \frac{2|x|}{\sigma^2 T} \tag{51}$$

Let $h_U$ denote the differential entropy of $X_{i1}(T_1)$ and $X_{i2}(T_1)$ i.e. $h_U \triangleq h(X_{i1}(T_1)) = h(X_{i2}(T_1))$, then the lower bound on update rate is given by

$$R(\epsilon^2) \geq \max\left(2h_U - \log\left(\pi e\epsilon^2\right), 0\right) \text{ bits/packet} \tag{52}$$

$$U(\epsilon^2) \geq \frac{2}{T}\max\left(2h_U - \log\left(\pi e\epsilon^2\right), 0\right) \text{ bits/second} \tag{53}$$

*3) Exponential distribution of packet inter-arrival time:*  For the exponential packet arrival process, the probability distribution functions of $X_{i1}(T_1)$ and $X_{i2}(T_1)$ are given by

$$f_{X_1}(x) = f_{X_2}(x) = \int_0^\infty \frac{\alpha}{\sqrt{2\pi\sigma^2\tau}}e^{-\left(\frac{x^2}{2\sigma^2\tau} + \alpha\tau\right)}d\tau \tag{54}$$

Let $h_E$ denote the differential entropy of $X_{i1}(T_1)$ and $X_{i2}(T_1)$ i.e. $h_E \triangleq h(X_{i1}(T_1)) = h(X_{i2}(T_1))$, then the lower bound on update rate is given by

$$R(\epsilon^2) \geq \max\left(2h_E - \log\left(\pi e\epsilon^2\right), 0\right) \text{ bits/packet} \tag{55}$$

$$U(\epsilon^2) \geq \alpha\max\left(2h_E - \log\left(\pi e\epsilon^2\right), 0\right) \text{ bits/second} \tag{56}$$

*4) Comparison of update rates for various inter-arrival processes:*  Figures 1 and 2 show the plot of the lower bound on $U(\epsilon^2)$ against $\sigma^2$ and $E[S]$. It is observed that for high $\sigma^2$ and low $E[S]$, the rate at which a source must update its location servers becomes very high. Also it is observed that the rate required for deterministic packet arrival is higher than that required for uniform and exponential arrival processes. In fact the update rate for deterministic packet arrival process is higher than any other packet arrival process with the same mean inter-arrival time. Consider a packet arrival process with pdf $f_S(t)$ and mean $E[S]$, then $\text{Var}(X_{i1}(T_1))$ and $\text{Var}(X_{i2}(T_1))$ are given by

$$\int_0^\infty \int_{-\infty}^\infty x^2 \frac{1}{\sqrt{\pi \sigma^2 t}} e^{-\frac{x^2}{\sigma^2 t}} \mathrm{d}x f_S(t) \mathrm{d}t = \int_0^\infty \frac{\sigma^2 t}{2} f_S(t) \mathrm{d}t = \frac{\sigma^2 E[S]}{2} \tag{57}$$

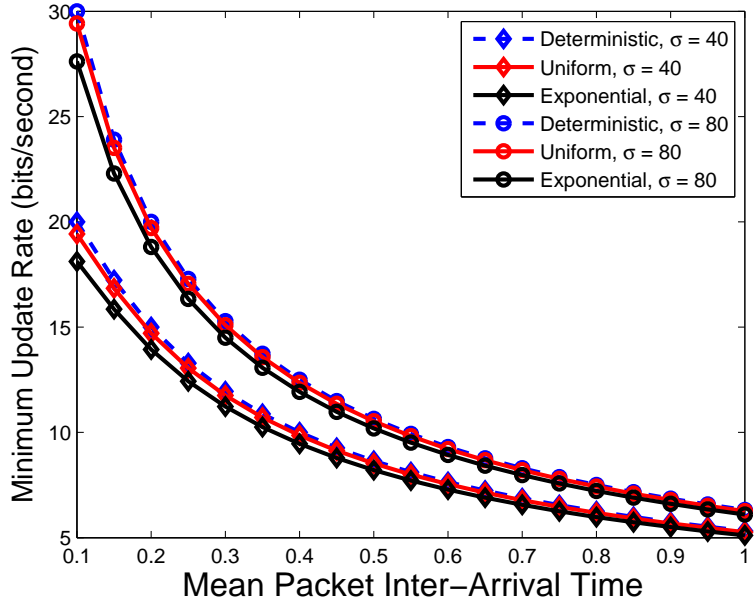

Fig. 2.   Update rate vs. mean packet inter-arrival time.

This implies that notwithstanding the packet arrival process, the variance of the change in location between two packet arrival instances depends only on $\sigma$ and $E[S]$. For the deterministic packet arrival process, $X_{i1}(T_1)$ and $X_{i2}(T_1)$ are Gaussian random variables (48). Since among random variables with the same variance Gaussian random variable has the highest entropy, deterministic packet arrival leads to the highest update rate.

## V. BEACON OVERHEAD

In this section we evaluate a lower bound on the minimum rate at which the nodes need to transmit beacons so that neighbors maintain a consistent neighborhood information. We then bound overhead in terms of bits per second. We first consider Brownian motion in one dimension and later extend the results to two dimension case.

### A. Notation and minimum beacon rate formulation

All the notations defined in subsections III and IV-A are used throughout this section. In this subsection we define some additional notations for the analysis.

*Definition 8:* $N_i(t)$ is the set of nodes that belong to the neighborhood of node $i$. That is

$$N_i(t) = \{j : |X_i(t) - X_j(t)| \le r, 1 \le j \le n, j \ne i\} \tag{58}$$

*Definition 9:* $\hat{N}_i(t)$ is the set of nodes that the node $i$ *perceives* to be its neighbors.

The set $\hat{N}_i(t)$ is constructed by node $i$ based on the beacons it receives. A node that belongs to $\hat{N}_i(t)$ may be excluded from $\hat{N}_i(t+\tau)$ if sufficient beacons are not received from the node during time interval $[t, t+\tau]$. Similarly, a node not belonging to $\hat{N}_i(t)$ may be included in $\hat{N}_i(t+\tau)$ if sufficient beacons are received from the node during time interval $[t, t+\tau]$. The deviation of $\hat{N}_i(t)$ from $N_i(t)$ depends on the rate at which the nodes transmit beacons.

*Definition 10:* $Z_{ij}(t)$ and $\hat{Z}_{ij}(t)$ $(1 \leq i, j \leq n, i \neq j)$ are indicator random variables, defined in the following manner

$$Z_{ij}(t) = \begin{cases} 1, & \text{if } j \in N_i(t) \\ 0, & \text{otherwise} \end{cases} \tag{59}$$

and

$$\hat{Z}_{ij}(t) = \begin{cases} 1, & \text{if } j \in \hat{N}_i(t) \\ 0, & \text{otherwise} \end{cases} \tag{60}$$

In other words, $Z_{ij}(t)$ equals 1 if node $j$ belongs to the neighborhood of node $i$ at time $t$. Note that $Z_{ij}(t)$ is a symmetric relation, i.e. $Z_{ij}(t) = Z_{ji}(t)$. On the other hand, $\hat{Z}_{ij}(t)$ is 1 if node node $i$ perceives node $j$ to be its neighbor at time $t$. Unlike $Z_{ij}(t)$, $\hat{Z}_{ij}(t)$ is not symmetric. That is $\hat{Z}_{ij}(t)$ may be 0 although $\hat{Z}_{ji}(t)$ is 1. This may happen because beacon transmission rate of $i$ is high enough to allow $j$ to maintain consistent neighborhood set while beacon transmission rate of node $j$ is not high enough to allow node $i$ to maintain consistent neighborhood set. The deviation of the perceived neighborhood, $\hat{N}_i(t)$, from the actual neighborhood, $N_i(t)$, is reflected by the deviation of $\hat{Z}_{ij}(t)$ from $Z_{ij}(t)$.

*Definition 11:* The difference of $\hat{Z}_{ij}(t)$ and $Z_{ij}(t)$ is defined as $E_{ij}(t)$, i.e.

$$E_{ij}(t) = Z_{ij}(t) - \hat{Z}_{ij}(t) \tag{61}$$

$E_{ij}(t) = 0$ implies that node $i$ has accurate information about whether $j$ belongs to its neighborhood or not. It is not necessary that $E_{ij}(t) = 0$ for all $t$, however it is desirable that $E_{ij}(t) = 0$ with high probability at all time instances when node $i$ has a packet to forward. This is because correct neighborhood information is highly critical for node $i$ to make correct forwarding decisions.

We can now state the minimum beacon rate problem in the following manner.

*Minimum beacon rate problem*: What is the minimum rate at which node $j$ must transmit beacons such that

$$P[E_{ij}(\tau_i(k)) = 0] \geq 1 - \delta \ \forall \ 1 \leq i \neq j \leq n, 1 \leq k < \infty \tag{62}$$

In order to formulate the above minimum beacon rate problem as a rate distortion problem we present two more definitions.

*Definition 12:* Let the vectors $Z_{ij}^N$ and $\hat{Z}_{ij}^N$ be defined in the following manner

$$Z_{ij}^N \triangleq \{Z_{ij}(\tau_i(1)), Z_{ij}(\tau_i(2)), \ldots, Z_{ij}(\tau_i(N))\} \tag{63}$$

$$\hat{Z}_{ij}^N \triangleq \{\hat{Z}_{ij}(\tau_i(1)), \hat{Z}_{ij}(\tau_i(2)), \ldots, \hat{Z}_{ij}(\tau_i(N))\} \tag{64}$$

*Definition 13:* Let $\mathcal{P}_N^{(b)}(\delta)$ denote the family of joint probability distribution function of $Z_{ij}^N$ and $\hat{Z}_{ij}^N$ such that $P[E_{ij}(\tau_i(k)) = 0] \leq 1 - \delta \ \forall \ 1 \leq k \leq N$.

The superscript in $\mathcal{P}_N^{(b)}(\delta)$ is used in order to distinguish the notation from the one used in the previous section. This superscript will be used for similar purpose in the rest of this section.

Thus the minimum beacon rate, $R^{(b)}(\delta)$, may be expressed in the following manner

$$R^{(b)}(\delta) = \lim_{N \to \infty} \min R_N^{(b)}(\delta) \tag{65}$$

where

$$R_N^{(b)}(\delta) = \min_{P_N \in \mathcal{P}_N^{(b)}(\delta)} \frac{1}{N} I_{P_N}(Z_{ij}^N; \hat{Z}_{ij}^N) \tag{66}$$

and $I_{P_N}(Z_{ij}^N; \hat{Z}_{ij}^N)$ is the mutual information between $Z_{ij}^N$ and $\hat{Z}_{ij}^N$. In the next subsection we evaluate a lower bound on $R^{(b)}(\delta)$.

### B. Beacon Rate Analysis for One-Dimension Networks

We first present the lower bound on the beacon rate of a node for the one-dimensional case and then extend the result to the two dimension case.

*Lemma 3:* The minimum beacon rate of node $j$, $R^{(b)}(\delta)$ is greater than equal to $R_1^{(b)}(\delta)$, that is

$$R^{(b)}(\delta) \geq R_1^{(b)}(\delta) \tag{67}$$

The proof of the above Lemma is similar to that of Lemma 1.

Following the result of Lemma 3, we need to find a lower bound on $R_1^{(b)}(\delta)$ in order to bound $R^{(b)}(\delta)$. We first find a lower bound on $R_1^{(b)}$ for the case when $Z_{ij}(0) = 1$ and then find the bound for the case when $Z_{ij}(0) = 0$. The bound for the these two cases provide answer to the following question: (i) If $i$ and $j$ are neighbors at time $t = 0$, then at what rate must $j$ transmit beacons such that node $i$ knows whether $j$ is neighbor of $i$ or not when it has a packet to send at $t = \tau_i(1)$; (ii) If $i$ and $j$ are not neighbors at $t = 0$, then at what rate must $j$ transmit beacons such that $i$ knows whether $j$ is a neighbor of $i$ or not when it has a packet to send at $t = \tau_i(1)$. Since the constraint $P[E_{ij}(\tau_i(1)) = 0] \geq 1 - \delta$ has to be satisfied for all $1 \leq i \leq n$, whether $i$ is a neighbor of $j$ or not at $t = 0$, the lower bound is maximum of beacon rate for the two cases.

*Lemma 4:* If $Z_{ij}(0) = 1$, then $R_1^{(b)}(\delta)$ is bounded by

$$R_1^{(b)}(\delta) \geq \max_{X_j(0) \in L_i} H(Z_{ij}(\tau_i(1))) - \mathcal{H}\left(\frac{\delta}{2}, 1 - \delta, \frac{\delta}{2}\right) \tag{68}$$

where

$$\mathcal{H}\left(\frac{\delta}{2}, \delta, \frac{\delta}{2}\right) \triangleq -\delta \log\left(\frac{\delta}{2}\right) - (1 - \delta) \log(1 - \delta) \tag{69}$$

and

$$L_i \triangleq [X_i(0) - r, X_i(0) + r] \tag{70}$$

i.e. $L_i$ is the set of possible positions of $j$ at time $t = 0$ such that $Z_{ij}(0) = 1$.

*Proof:* Recall that $R_1^{(b)}(\delta)$ is given by

$$R_1^{(b)}(\delta) = \inf_{P_1 \in \mathcal{P}_1(\delta)} I_{P_1}(Z_{ij}(\tau_1); \hat{Z}_{ij}(\tau_1)) \tag{71}$$

Now $I_{P_1}(Z_{ij}(\tau_1); \hat{Z}_{ij}(\tau_1))$ is given by

$$\begin{aligned}
I_{P_1}(Z_{ij}(\tau_1); \hat{Z}_{ij}(\tau_1)) &= H(Z_{ij}(\tau_1)) - H(Z_{ij}(\tau_1)|\hat{Z}_{ij}(\tau_1)) \tag{72}\\
&= H(Z_{ij}(\tau_1)) - H(Z_{ij}(\tau_1) - \hat{Z}_{ij}(\tau_1)|\hat{Z}_{ij}(\tau_1)) \tag{73}\\
&\geq H(Z_{ij}(\tau_1)) - H(Z_{ij}(\tau_1) - \hat{Z}_{ij}(\tau_1)) \tag{74}\\
&= H(Z_{ij}(\tau_1)) - H(E_{ij}(\tau_1)) \tag{75}
\end{aligned}$$

We know that the probability distribution of $E_{ij}(\tau_1)$ is given by

$$E_{ij}(\tau_1)) = \begin{cases} -1, & \text{w.p. } p_1 \\ 0, & \text{w.p. } p_2 \\ 1, & \text{w.p. } p_3 \end{cases} \tag{76}$$

where $p_2 \geq 1 - \delta$, $p_1 + p_3 \leq \delta$ and $p_1 + p_2 + p_3 = 1$ (since $P_1 \in \mathcal{P}_1(\delta)$). Under these constraints $H(E_{ij}(\tau_1)))$ is maximized when $p_2 = 1 - \delta$ and $p_1 = p_3 = \delta/2$, when $H(E_{ij}(\tau_1)) = \mathcal{H}\left(\frac{\delta}{2}, 1 - \delta, \frac{\delta}{2}\right)$. Thus

$$I_{P_1}(Z_{ij}(\tau_1); \hat{Z}_{ij}(\tau_1)) \geq H(Z_{ij}(\tau_1)) - \mathcal{H}\left(\frac{\delta}{2}, 1 - \delta, \frac{\delta}{2}\right) \tag{77}$$

Now $H(Z_{ij}(\tau_1))$ depends on the position of node $j$ at $t = 0$, $X_j(0)$. We know that $j \in N_i(0)$ which implies that $X_j(0) \in L_i$. Depending upon $f_\tau(t)$, $H(Z_{ij}(\tau_1))$ is maximized for some $X_j(0) = x \in L_i$. In order to ensure that $P[E_{ij}(\tau_i(k)) = 0] \leq 1 - \delta$ $\forall$ $i$, the beacon rate must take care of this worst case. Thus

$$I_{P_1}(Z_{ij}(\tau_1); \hat{Z}_{ij}(\tau_1)) \geq \max_{X_j(0) \in L_i} H(Z_{ij}(\tau_i(1))) - \mathcal{H}\left(\frac{\delta}{2}, 1 - \delta, \frac{\delta}{2}\right)$$

¿From the above equation and (71) we get (68).                                                                                                            ∎

We now outline how to evaluate $H(Z_{ij}(\tau_i(1)))$ Without loss of generality we may assume that $X_i(0) = 0$. So if $Z_{ij}(0) = 1$, then $X_j(0) = l$ where $-r \leq l \leq r$. ¿From the point of reference of node $i$, node $j$ performs Brownian motion with variance $2\sigma^2$. So

$$P[Z_{ij}(\tau_i(1)) = 1 | X_j(0) = l \in L_i, \tau_i(1) = \tau'] = \frac{1}{2}\text{erf}\left(\frac{r-l}{\sqrt{4\sigma^2\tau'}}\right) + \frac{1}{2}\text{erf}\left(\frac{r+l}{\sqrt{4\sigma^2\tau'}}\right) \tag{78}$$

Now using the fact that the inter-arrival time of packets to be served by $i$ has distribution $f_\tau(t)$, we get

$$P[Z_{ij}(\tau_i(1)) = 1 | X_j(0) = l \in L_i] = \frac{1}{2}\int_{t=0}^{\infty}\text{erf}\left(\frac{r-l}{\sqrt{4\sigma^2 t}}\right)f_\tau(t)dt + \frac{1}{2}\int_{t=0}^{\infty}\text{erf}\left(\frac{r+l}{\sqrt{4\sigma^2 t}}\right)f_\tau(t)dt \tag{79}$$

We know that $H(Z_{ij}(\tau_i(1))) = \mathcal{H}(P[Z_{ij}(\tau_i(1)) = 1])$, where $\mathcal{H}(x) = -x\log(x) - (1-x)\log(1-x)$ $(0 \leq x \leq 1)$. We know that $\mathcal{H}(x)$ is maximum at $x = 0.5$, symmetric about $x = 0.5$ and is strictly increasing and decreasing in the interval $[0, 0.5)$ and $(0.5, 1]$ respectively.

*Definition 14:* Let $l^\star$ be defined as

$$l^\star \triangleq \arg\min_{-r \leq l \leq r} |P[Z_{ij}(\tau_i(1)) = 1 | X_j(0) = l] - 0.5| \tag{80}$$

In other words $l^\star$ is the value of $X_j(0)$ for which $H(Z_{ij}(\tau_i(1)))$ is maximized. This leads us to the following Corollary.

*Corollary 1:* The minimum beacon transmission rate of node $j$, denoted by $R^{(b_1)}(\delta)$, such that with probability at least $1-\delta$ the current neighbors of $j$ know whether $j$ belongs to their neighborhood at the time of forwarding next packet equals

$$R^{(b_1)}(\delta) \geq \mathcal{H}(p(l^\star)) - \mathcal{H}\left(\frac{\delta}{2}, 1-\delta, \frac{\delta}{2}\right) \text{ beacons/msg} \tag{81}$$

where $p(l^\star) = P[Z_{ij}(\tau_i(1)) = 1 | |X_j(0) - X_i(0)| = l^\star]$ and $l^\star$ is given by (80).

We now turn our attention to the second question i.e. at what rate should $j$ transmit such that with probability at least $1-\delta$ all the nodes that are not neighbors of $j$ at $t = 0$ know whether $j$ belongs to their neighborhood or not when they have a packet to forward?

It is easy to see that Lemma 4 holds even if $Z_{ij}(0) = 0$. We formally state a similar Lemma without proof for the case when $Z_{ij}(0) = 0$.

*Lemma 5:* If $Z_{ij}(0) = 0$, then $R_1^{(b)}(\delta)$ is bounded by

$$R_1^{(b)} \geq \max_{X_j(0) \in L_i'} H(Z_{ij}(\tau_i(1))) - \mathcal{H}\left(\frac{\delta}{2}, 1-\delta, \frac{\delta}{2}\right) \tag{82}$$

where $\mathcal{H}\left(\frac{\delta}{2}, 1-\delta, \frac{\delta}{2}\right)$ is given by (69) and $L_i'$ is the set of possible positions of node $j$ at $t = 0$ such that $Z_{ij} = 0$, i.e.,

$$L_i' \triangleq \{x : |x - X_i(0)| \geq r\} \tag{83}$$

Again, without loss of generality we assume that $X_i(0) = 0$. Let $X_j(0) = l$, such that $|l| \geq r$, i.e. $j$ does not belong to the neighborhood of $i$ at time $t = 0$. $P[Z_{ij}(\tau_i(1)) = 1 | X_j(0) = l, |l| \geq r, \tau_i(1) = \tau']$ is given by

$$P[Z_{ij}(\tau_i(1)) = 1 | X_j(0) = l, |l| \geq r, \tau_i(1) = \tau'] = \frac{1}{2}\text{erf}\left(\frac{|l|+r}{\sqrt{4\sigma^2\tau'}}\right) - \frac{1}{2}\text{erf}\left(\frac{|l|-r}{\sqrt{4\sigma^2\tau'}}\right) \tag{84}$$

Thus

$$P[Z_{ij}(\tau_i(1)) = 1 | X_j(0) = l, |l| \geq r] = \frac{1}{2}\int_0^\infty \text{erf}\left(\frac{|l|+r}{\sqrt{4\sigma^2 t}}\right)f_\tau(t)dt - \frac{1}{2}\int_0^\infty \text{erf}\left(\frac{|l|-r}{\sqrt{4\sigma^2 t}}\right)f_\tau(t)dt \tag{85}$$

Note that $P[Z_{ij}(\tau_i(1)) = 1 | X_j(0) = l, |l| \geq r] \leq 0.5$ and its value is maximized for $|l| = r$. Since $\mathcal{H}(x)$ is an increasing function of $x$ in the interval $[0, 0.5)$, the $|l| = r$ maximizes the value of $H(Z_{ij}(\tau_i(1))) = \mathcal{H}(P[Z_{ij}(\tau_i(1)) = 1 | X_j(0) = l, |l| \geq r])$. This leads to the following Corollary.

*Corollary 2:* The minimum beacon transmission rate of node $j$, denoted by $R^{(b_2)}(\delta)$, such that with probability at least $1-\delta$ the nodes that are not current neighbors of $j$ know whether $j$ belongs to their neighborhood at time of forwarding the

next packet equals

$$R^{(b_2)}(\delta) \geq \mathcal{H}(p(r)) - \mathcal{H}\left(\frac{\delta}{2}, 1 - \delta, \frac{\delta}{2}\right) \text{ beacons/msg} \tag{86}$$

where

$$p(r) = \frac{1}{2} \int_0^\infty \text{erf}\left(\frac{r}{\sqrt{\sigma^2 t}}\right) f_\tau(t) dt$$

To summarize, Corollary 1 provides a lower bound on the beacon transmission rate such that each of the current neighbors are able to maintain consistent neighborhood information with high probability. Corollary 2 provides a lower bound on the beacon transmission rate of a node such that all nodes that are not neighbor of the node know with high probability if the node joins their neighborhood. The minimum beacon transmission rate is therefore the maximum of the two rates given by (81) and (86).

However note that according to the definition of $l^\star$, $\mathcal{H}(p(l^\star)) \geq \mathcal{H}(p(r))$. This implies that the lower bound on $R^{(b_2)}(\delta)$ in (86) is always less than the lower bound on $R^{(b_1)}(\delta)$ in (81). Thus the lower bound on the minimum beacon transmission rate is given by the lower bound on $R^{(b_1)}(\delta)$ in (81). The following theorem formally states this discussion.

*Theorem 3:* The lower bound on the minimum beacon transmission rate of a node such that the constraint in equation (62) is satisfied is given

$$R^{(b)}(\delta) \quad \geq \quad \max\left(R^{(b_1)}(\delta), R^{(b_2)}(\delta)\right) \tag{87}$$

$$\geq \quad \mathcal{H}(p(l^\star)) - \mathcal{H}\left(\frac{\delta}{2}, 1 - \delta, \frac{\delta}{2}\right) \text{ beacons/msg} \tag{88}$$

where $R^{(b_1)}(\delta)$, $R^{(b_2)}(\delta)$ and $l^\star$ are given by (81), (86) and (80) respectively.

When $\mathcal{H}(p(l^\star)) < \mathcal{H}\left(\frac{\delta}{2}, 1 - \delta, \frac{\delta}{2}\right)$, left hand side of (88) will be meaningless. Thus (88) may be expressed as

$$R^{(b)}(\delta) \geq \max\left(\mathcal{H}(p(l^\star)) - \mathcal{H}\left(\frac{\delta}{2}, 1 - \delta, \frac{\delta}{2}\right), 0\right) \text{ beacons/msg} \tag{89}$$

The minimum overhead in bits/second, denoted by $U^{(b)}(\delta)$, is given by

$$U^{(b)}(\delta) \geq \frac{B}{E[\tau]} \max\left(\mathcal{H}(p(l^\star)) - \mathcal{H}\left(\frac{\delta}{2}, 1 - \delta, \frac{\delta}{2}\right), 0\right) \text{ bits/sec} \tag{90}$$

where $E[\tau]$ is the expected packet inter-arrival time and $B$ is the size of beacon packet in bits.

We now consider various packet inter-arrival distributions, $f_\tau(t)$ in order to obtain more insights into the results obtained in this section.

*1) Deterministic inter-arrival time :* Consider deterministic inter-arrival time where at each node a packet to be forwarded arrives every $T$ seconds. For this case $f_\tau(t)$ is given by

$$f_\tau(t) = \delta(t - T)$$

So $p(l) \triangleq P[Z_{ij}(\tau_i(1)) = 1 | X_j(0) = l \in L_i]$ is given by

$$p(l) = P[Z_{ij}(\tau_i(1)) = 1 | X_j(0) = l \in L_i] = \frac{1}{2} \int_0^\infty \text{erf}\left(\frac{r - l}{\sqrt{4\sigma^2 t}}\right) \delta(t - T) dt + \frac{1}{2} \int_0^\infty \text{erf}\left(\frac{r + l}{\sqrt{4\sigma^2 t}}\right) \delta(t - T) dt$$
$$= \frac{1}{2} \left(\text{erf}\left(\frac{r - l}{\sqrt{4\sigma^2 T}}\right) + \text{erf}\left(\frac{r + l}{\sqrt{4\sigma^2 T}}\right)\right) \tag{91}$$

In order to find $R^{(b)}(\delta)$, we need to find $l^\star$ as defined in (80). Note that $p(l)$ is decreases monotonically as $l$ increases form 0 to $r$. Thus $l^\star$ may take the following values: (i) If $p(r) \geq 0.5$, then $l^\star = r$; (ii) If $p(0) \leq 0.5$, then $l^\star = 0$; and (iii) If $p(0) > 0.5$ and $p(r) < 0.5$, then $0 < l^\star < r$ and $\mathcal{H}(p(l^\star)) = 1$. Thus we have the following relation

$$R^{(b)}(\delta) \geq \begin{cases} \max\left(\mathcal{H}\left(\frac{1}{2}\mathrm{erf}\left(\frac{r}{\sqrt{\sigma^2 T}}\right)\right) - \mathcal{H}\left(\frac{\delta}{2}, 1 - \delta, \frac{\delta}{2}\right), 0\right), & \text{if } p(r) \geq 0.5 \\ \max\left(\mathcal{H}\left(\mathrm{erf}\left(\frac{r}{\sqrt{4\sigma^2 T}}\right)\right) - \mathcal{H}\left(\frac{\delta}{2}, 1 - \delta, \frac{\delta}{2}\right), 0\right), & \text{if } p(0) \leq 0.5 \\ \max\left(1 - \mathcal{H}\left(\frac{\delta}{2}, 1 - \delta, \frac{\delta}{2}\right), 0\right), & \text{if } p(r) < 0.5 < p(0) \end{cases} \quad (92)$$

The above relation may be expressed in a more compact manner in the following manner

$$R^{(b_1)}(\delta) \geq \max\left(\mathcal{H}\left(\frac{1}{2}\mathrm{erf}\left(\frac{r}{\sqrt{\sigma^2 T}}\right)\right)\mathcal{U}(p(r) - 0.5) + \mathcal{H}\left(\mathrm{erf}\left(\frac{r}{\sqrt{4\sigma^2 T}}\right)\right)\mathcal{U}(0.5 - p(r)) + 1 \cdot \mathcal{U}(0.5 - p(r))\mathcal{U}(p(0) - 0.5)\right.$$
$$\left. - \mathcal{H}\left(\frac{\delta}{2}, 1 - \delta, \frac{\delta}{2}\right), 0\right) \text{ beacons/msg} \quad (93)$$

where $\mathcal{U}(x)$ is the unit step function, defined by

$$\mathcal{U}(x) = \begin{cases} 1, & x \geq 0 \\ 0, & \text{otherwise} \end{cases} \quad (94)$$

Let $B$ represent the size of a beacon packet in bits, then the overhead in bits/second is given by

$$U^{(b)}(\delta) \geq \frac{B}{T}\max\left(\mathcal{H}\left(\frac{1}{2}\mathrm{erf}\left(\frac{r}{\sqrt{\sigma^2 T}}\right)\right)\mathcal{U}(p(r) - 0.5) + \mathcal{H}\left(\mathrm{erf}\left(\frac{r}{\sqrt{4\sigma^2 T}}\right)\right)\mathcal{U}(0.5 - p(r))\right.$$
$$\left. + 1 \cdot \mathcal{U}(0.5 - p(r))\mathcal{U}(p(0) - 0.5) - \mathcal{H}\left(\frac{\delta}{2}, 1 - \delta, \frac{\delta}{2}\right), 0\right) \text{ bits/sec} \quad (95)$$

*2) Uniform inter-arrival time distribution :* Now suppose that the packet inter-arrival time has a uniform distribution given by

$$f_\tau(t) = \begin{cases} \frac{1}{T}, & 0 \leq t \leq T \\ 0, & \text{otherwise} \end{cases} \quad (96)$$

Thus $p(l) \triangleq P[Z_{ij}(\tau_i(1))|X_j(0) = l \in L_i]$ is given by

$$p(l) = P[Z_{ij}(\tau_i(1))1|X_j(0) = l \in L_i] = \frac{1}{2T}\int_0^T \mathrm{erf}\left(\frac{r - l}{\sqrt{4\sigma^2 t}}\right)dt + \frac{1}{2T}\int_0^T \mathrm{erf}\left(\frac{r + l}{\sqrt{4\sigma^2 t}}\right)dt \quad (97)$$

Evaluating the first integral in the above equation we get

$$\frac{1}{2T}\int_0^T \mathrm{erf}\left(\frac{r - l}{\sqrt{4\sigma^2 t}}\right)dt = \frac{(r - l)^2}{4\sigma^2 T}\mathrm{erf}\left(\frac{r - l}{\sqrt{4\sigma^2 T}}\right) + \frac{r - l}{\sqrt{4\pi\sigma^2 T}}\exp\left(-\frac{(r - l)^2}{4\sigma^2 T}\right) + \frac{1}{2}\mathrm{erf}\left(\frac{r - l}{\sqrt{4\sigma^2 T}}\right) - \frac{(r - l)^2}{4\sigma^2 T} \quad (98)$$

Similarly the second integral yields

$$\frac{1}{2T}\int_0^T \mathrm{erf}\left(\frac{r + l}{\sqrt{4\sigma^2 t}}\right)dt = \frac{(r + l)^2}{4\sigma^2 T}\mathrm{erf}\left(\frac{r + l}{\sqrt{4\sigma^2 T}}\right) + \frac{r + l}{\sqrt{4\pi\sigma^2 T}}\exp\left(-\frac{(r + l)^2}{4\sigma^2 T}\right) + \frac{1}{2}\mathrm{erf}\left(\frac{r + l}{\sqrt{4\sigma^2 T}}\right) - \frac{(r + l)^2}{4\sigma^2 T} \quad (99)$$

Thus $p(l)$ is given by

$$p(l) = \frac{(r - l)^2}{4\sigma^2 T}\mathrm{erf}\left(\frac{r - l}{\sqrt{4\sigma^2 T}}\right) + \frac{(r + l)^2}{4\sigma^2 T}\mathrm{erf}\left(\frac{r + l}{\sqrt{4\sigma^2 T}}\right) + \frac{r - l}{\sqrt{4\pi\sigma^2 T}}\exp\left(-\frac{(r - l)^2}{4\sigma^2 T}\right) + \frac{r + l}{\sqrt{4\pi\sigma^2 T}}\exp\left(-\frac{(r + l)^2}{4\sigma^2 T}\right)$$
$$+ \frac{1}{2}\left(\mathrm{erf}\left(\frac{r - l}{\sqrt{4\sigma^2 T}}\right) + \mathrm{erf}\left(\frac{r + l}{\sqrt{4\sigma^2 T}}\right)\right) - \frac{r^2 + l^2}{2\sigma^2 T} \quad (100)$$

Since $\mathrm{erf}\left(\frac{r - l}{\sqrt{4\sigma^2 t}}\right) + \mathrm{erf}\left(\frac{r - l}{\sqrt{4\sigma^2 t}}\right)$ strictly decreases with $l$, $p(l)$ is also a strictly decreasing function of $l$. Thus, similar to the deterministic arrival case, $l^\star$ may take the following values: (i) If $p(r) \geq 0.5$, then $l^\star = r$; (ii) If $p(0) \leq 0.5$, then $l^\star = 0$;

and (iii) If $p(0) > 0.5$ and $p(r) < 0.5$, then $0 < l^\star < r$ and $\mathcal{H}(p(l^\star)) = 1$. Thus

$$p(l^\star) = \begin{cases} \frac{r^2}{\sigma^2 T}\mathrm{erf}\left(\frac{r}{\sqrt{\sigma^2 T}}\right) + \frac{r}{\sqrt{\pi}\sigma^2 T}\exp\left(-\frac{r^2}{\sigma^2 T}\right) \\ \quad + \frac{1}{2}\mathrm{erf}\left(\frac{r}{\sqrt{\sigma^2 T}}\right) - \frac{r^2}{\sigma^2 T}, & \text{if } p(r) \geq 0.5 \\ \frac{r^2}{2\sigma^2 T}\mathrm{erf}\left(\frac{r}{\sqrt{4\sigma^2 T}}\right) + \frac{r}{\sqrt{\pi}\sigma^2 T}\exp\left(-\frac{r^2}{4\sigma^2 T}\right) \\ \quad + \mathrm{erf}\left(\frac{r}{\sqrt{4\sigma^2 T}}\right) - \frac{r^2}{2\sigma^2 T}, & \text{if } p(0) \leq 0.5 \\ 0.5, & \text{if } p(0) > 0.5 > p(r) \end{cases} \tag{101}$$

Substituting the above expression of $p(l^\star)$ in the expression for the lower bound on the beacon update rate for the case of uniform inter-arrival time distribution may be calculated using (89). For uniform distribution of inter-arrival time, the lower bound on the overhead caused by the beacon updates is thus by

$$U^{(b)}(\delta) \geq \frac{2B}{T}\max\left(\mathcal{H}(p(l^\star)) - \mathcal{H}\left(\frac{\delta}{2}, 1 - \delta, \frac{\delta}{2}\right), 0\right) \text{ bits/sec} \tag{102}$$

where $B$ is the size of beacon packet and $p(l^\star)$ is given by (101).

*3) Exponential inter-arrival distribution :* Suppose that the inter-arrival packet time is exponentially distributed with mean $\frac{1}{\alpha}$ i.e.

$$f_\tau(t) = \alpha e^{-\alpha t} \tag{103}$$

Now for the exponential distribution $p(l)$ is given by

$$p(l) = \frac{\alpha}{2}\int_0^\infty \mathrm{erf}\left(\frac{r-l}{\sqrt{4\sigma^2 t}}\right)\exp(-\alpha t)dt + \frac{\alpha}{2}\int_0^\infty \mathrm{erf}\left(\frac{r+l}{\sqrt{4\sigma^2 t}}\right)\exp(-\alpha t)dt \tag{104}$$

Since $p(l)$ is a decreasing function of $l$, $p(l^\star)$ is given by

$$p(l^\star) = \begin{cases} p(r), & \text{if } p(r) \geq 0.5 \\ p(0), & \text{if } p(0) \leq 0.5 \\ 0.5, & \text{if } p(r) \geq 0.5 \end{cases} \tag{105}$$

and a lower bound on overhead caused by beacon updates is thus given by

$$U^{(b)}(\delta) = \alpha B \max\left(\mathcal{H}\left(p(l^\star)\right) - \mathcal{H}\left(\frac{\delta}{2}, 1 - \delta, \frac{\delta}{2}\right), 0\right) \tag{106}$$

It is not possible to find a closed form expression for the integrals in equation (104) and numerical methods may be used to evaluate the lower bound on $U^{(b)}(\delta)$ for the exponential inter-arrival time distribution.

## C. Beacon Rate Analysis for Two-Dimensional Networks

In this subsection we extend the minimum beacon rate analysis to two dimensional networks. For a arbitrary node pair $i$ and $j$, we choose an orthogonal coordinate system such that $X_{i1}(0) = X_{i2}(0) = 0$, $X_{j1}(0) = l$, and $X_{j2}(0) = 0$. That is, the origin of the coordinate system corresponds to the position of node $i$ at $t = 0$ and the x-axis of the coordinate system corresponds to the line joining the position of nodes $i$ and $j$ at $t = 0$. It can be easily verified that a Brownian motion with variance $\sigma^2$ can be decomposed into two independent Brownian motions with variance $\sigma^2/2$ along each axis. Also note that Lemmas 3 and 4 hold for the two dimensional case as well and may be proved in a similar manner. Thus the minimum beacon rate, $R^{(b)}(\delta)$, satisfies the following relationship

$$R^{(b)}(\delta) \geq H\left(Z_{ij}(\tau_i(1))\right) - \mathcal{H}\left(\frac{\delta}{2}, 1 - \delta, \frac{\delta}{2}\right) \tag{107}$$

Similar to the approach in the last section, we proceed by individually considering the cases $Z_{ij}(0) = 1$ and $Z_{ij}(0) = 0$.

For the case when $Z_{ij}(0) = 1$, the probability that $j$ is in the neighborhood of $i$ when $i$ has a packet to send $(p(l))$ is given

by

$$p(l) \triangleq P\left[Z_{ij}(\tau) = 1 | X_{j1}(0) = l, X_{j2}(0) = 0, |l| \leq r\right] = \int_{x=-r}^{x=r} P\left[X_{j1}(\tau_i(1)) = x | X_{j1}(0) = l\right] \cdot$$
$$P\left[-\sqrt{r^2 - x^2} \leq X_{j2}(\tau_i(1)) \leq \sqrt{r^2 - x^2} | X_{j2}(0) = 0\right]$$

Relative to node $i$, node $j$ performs Brownian motion with variance $2\sigma^2$. Thus

$$P\left[X_{j1}(\tau) = x | X_{j1}(\tau) = l\right] = \frac{1}{\sqrt{2\pi\sigma^2\tau}} \exp\left(-\frac{(l-x)^2}{2\sigma^2\tau}\right) dx$$

and

$$P\left[-\sqrt{r^2 - x^2} \leq X_{j2}(\tau) \leq \sqrt{r^2 - x^2} | X_{j2}(0) = 0\right] = \text{erf}\left(\frac{\sqrt{r^2 - x^2}}{\sqrt{2\sigma^2\tau}}\right)$$

Therefore,

$$p(l) = \int_0^\infty \int_{-r}^r \frac{1}{\sqrt{2\pi\sigma^2 t}} \exp\left(-\frac{(l-x)^2}{2\sigma^2 t}\right) \text{erf}\left(\frac{\sqrt{r^2 - x^2}}{\sqrt{2\sigma^2 t}}\right) f_\tau(t) dx\, dt \tag{108}$$

Thus in order to satisfy (62) at all neighbors that are neighbor at time 0, node $j$ must transmit beacon at a rate higher than

$$\mathcal{H}\left(p(l^\star)\right) - \mathcal{H}\left(\frac{\delta}{2}, 1 - \delta, \frac{\delta}{2}\right) \tag{109}$$

where $l^\star$ is given by (80).

Now consider the case when $j$ does not belong to the neighborhood of $i$ at $t = 0$. It can be easily verified that the probability that $Z_{ij}(\tau_i(1)) = 1$ given that $Z_{ij}(0) = 0$ ($p'(l)$) is given by the same expression as $p(l)$ in equation (108). $p'(l)$ increases with decrease in $|l|$ and is maximized for $|l| = r$. For other values of $l > r$, $p'(l) < 0.5$. Thus similar to the one-dimensional networks, the beacon transmission rate is determined by the rate required to satisfy (62) at the initial neighbors. This leads the following theorem.

*Theorem 4:* The lower bound on the minimum beacon transmission rate of a node such that the constraint in equation (62) is satisfied is given

$$R^{(b)}(\delta) \geq \max\left(\mathcal{H}\left(p(l^\star)\right) - \mathcal{H}\left(\frac{\delta}{2}, 1 - \delta, \frac{\delta}{2}\right), 0\right) \text{ beacons/pkt} \tag{110}$$

where $p(l)$ and $l^\star$ are given by (108) and (80) respectively. The beacon transmission overhead in bits per second, $U^{(b)}(\delta)$, is given by

$$U^{(b)}(\delta) \geq \frac{B}{E[\tau]} \max\left(\mathcal{H}(p(l^\star)) - \mathcal{H}\left(\frac{\delta}{2}, 1 - \delta, \frac{\delta}{2}\right), 0\right) \text{ bits/sec} \tag{111}$$

### D. Comparison of Beacon Transmission Rates for Various Arrival Processes

The closed form expression for the integral in (108) cannot be found. So we use numerical computations to evaluate $R^{(b)}(\delta)$ for deterministic, uniform and exponential packet arrival processes. Figures 3 and 4 show plots of minimum beacon rate in bits per second. Figure 3 shows the plot of minimum beacon transmission rate against variance of Brownian motion for different mean packet inter-arrival times. It is observed that for low variance the rate is almost constant, while as the the variance increases the rate starts decreasing. When the variance of Brownian motion is very small, the variance of the change in position of a node within a packet arrival epoch is also small. For this case, $l^\star = r$ and $p(l^\star) \approx 0.5$ which leads to high beacon rate. As the variance increases, the probability that two neighbors remain neighbors at the end of a packet arrival epoch is very small, no matter what the initial position of nodes might be. That is, when $\sigma^2$ is high, $p(l) < 0.5 \; \forall \; l$, which leads to low beacon rate when variance is high. This is illustrated in Figure 5. Suppose the two nodes shown in the figure have communication range of two units. Initially nodes 1 and 2 are located at $X = 0$ and $X = 2$ respectively. With respect to node 1, the pdf of position of node 2 when the next packet is forwarded is shown for $\sigma^2 = 10$ and $\sigma^2 = 80$. It is observed that for $\sigma^2 = 10$ the probability that nodes 2 will be in neighborhood of node 1 at the next packet arrival instant is approximately 0.5. This corresponds to large uncertainty which leads to high beacon update rate. On the other hand when $\sigma^2 = 80$ the probability that nods 1 and 2 are neighbors at the next packet arrival instant is less than $0.5$. It is therefore more likely that the two nodes will not be neighbors which reduces the uncertainty and hence leads to lower beacon update rate. The trend shown in Figure 3 implies that when nodes are highly mobile they need to transmit beacons less frequently and the membership of nodes in a neighborhood may be more efficiently deciphered by the absence of beacons. Figure 4 shows that as the rate of packet arrival increases, the beacon overhead may become prohibitively high. Also, for a given packet arrival rate, it is observed that the rate for a deterministic packet arrival process is smaller than that for exponential and uniform arrivals. This is because the probability that a node leaves the neighborhood of a certain neighbor within a packet inter-arrival duration is the highest ($p(l)$ is close to 1) for deterministic arrival. For the uniform and exponential distributions the probability that packet inter-arrival time is less than the mean inter-arrival time is $0.5$ and $0.63$ respectively. Thus the probability that a node moves out of neighborhood during an inter-arrival duration is smaller than that for the deterministic arrival process.

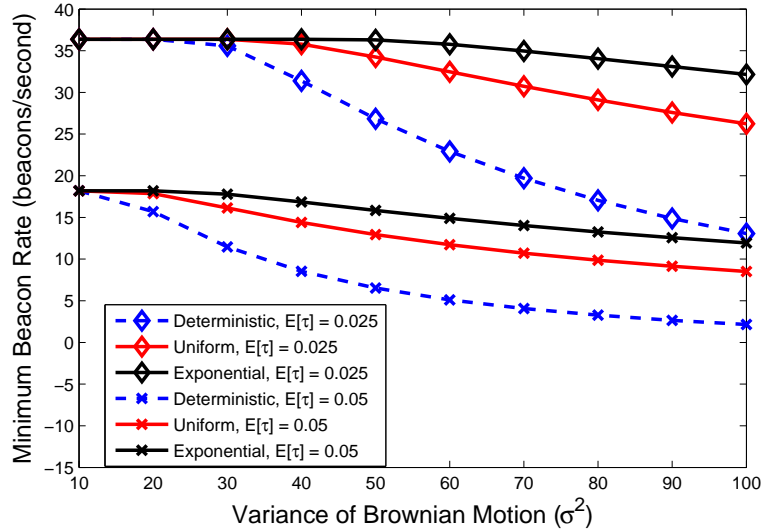

Fig. 3.   Beacons per second versus variance of Brownian motion.

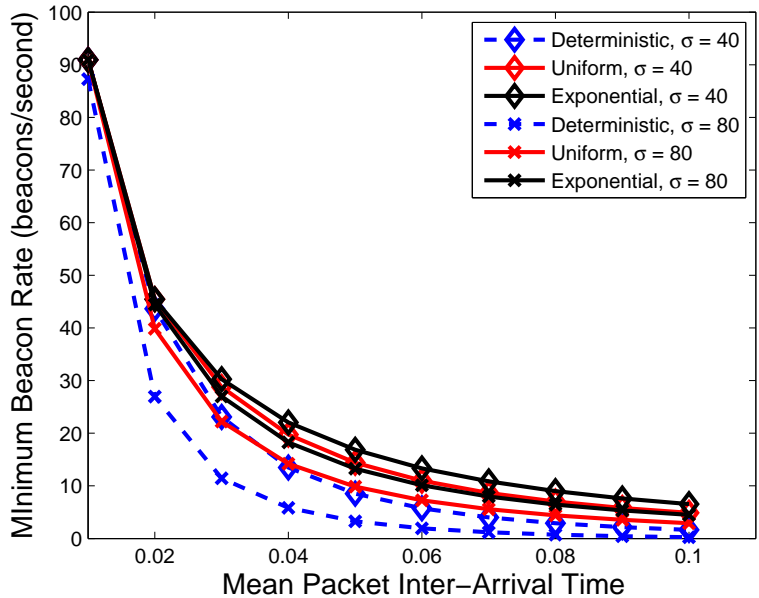

Fig. 4.   Beacons per second versus the mean inter-arrival time of packets to be forwarded at a node.

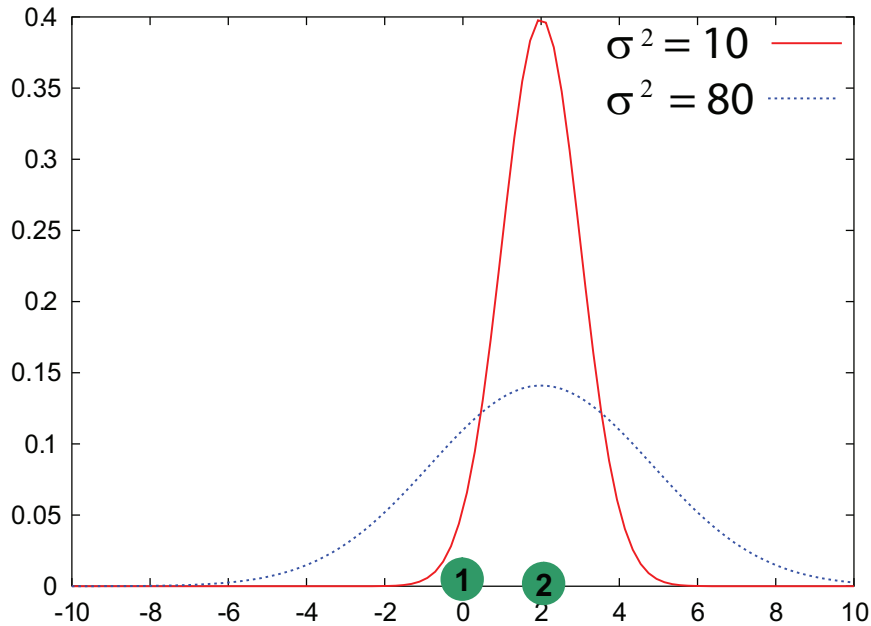

Fig. 5.   The pdf of position of node 2 with respect to node 1 given their initial positions.

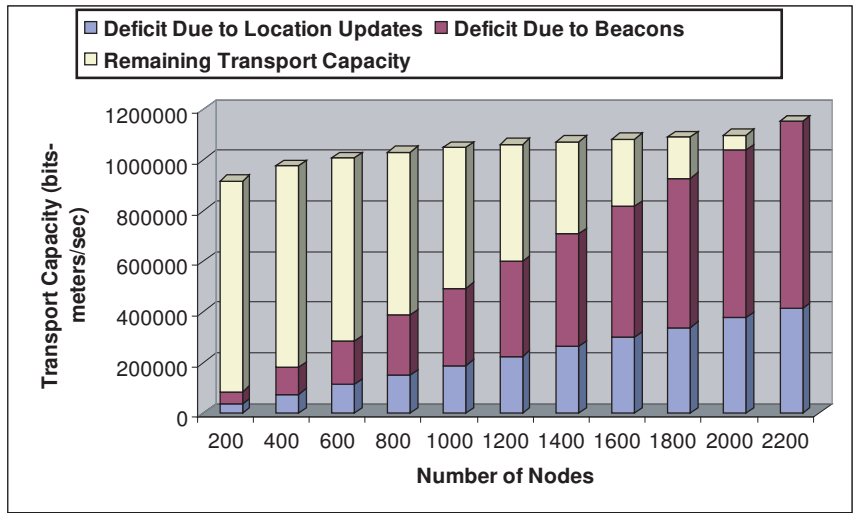

Fig. 6.   Capacity deficit versus number of nodes.

## VI. Capacity Deficit

A wireless ad hoc network is said to transport one bit-meter when a bit is transmitted over a distance of one meter [5]. The transport capacity of a network (in bit-meters per second) is defined as the supremum over the set of feasible rate vectors of the distance weighted sum of rates [27]. The transport capacity is expressed as $\lambda n \overline{L}$, where $\lambda$ is the average arrival rate at the nodes, $n$ is the number of nodes and $\overline{L}$ is the average distance traveled by the bits. It is shown in [5] that the transport capacity of an arbitrary wireless network is $\Theta\left(W\sqrt{nA}\right)$ where $n$, $W$ and $A$ are the number of nodes deployed, transmission rate of the nodes and area over which the network is deployed respectively. It is shown in [5] that for a particular interference model known as the *Protocol Model*, the upper bound on the transport capacity of an arbitrary wireless network is given by

$$\lambda n \overline{L} \leq \frac{\sqrt{8}}{\pi}\frac{1}{\Delta}W\sqrt{nA} \quad \text{bit} - \text{meters/second} \tag{112}$$

Let $\eta$ denote the expected distance between a node and its location server. Thus, on average, the location update information of a node travels at least $\eta$ meters before reaching its location server. Thus the average overhead incurred by a node for updating its location information is at least $\eta U(\epsilon^2)$ bit-meters/second and the overhead incurred by location update information on the network equals at least $n\eta U(\epsilon^2)$ bit-meters/second, where $U(\epsilon^2)$ is given by (47). A beacon transmitted by a node travels a distance equal to the communication radius. Thus the overhead incurred by the beacon packets on the network is at least $nrU^{(b)}(\delta)$, where $U^{(b)}(\delta)$ is given by (111). Thus the total transport capacity deficit due to the routing overhead is at least $n\eta U(\epsilon^2) + nrU^{(b)}(\delta)$ bit-meters/second. This leads to the following theorem.

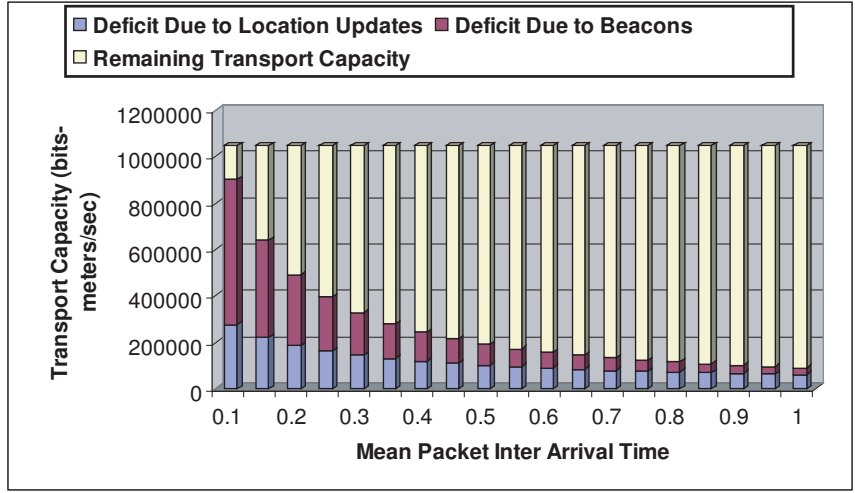

Fig. 7.   Capacity deficit versus packet inter arrival time.

*Theorem 5:* For the Protocol Model, the upper bound on the *residual transport capacity* available to an arbitrary network for transmitting data ($\lambda n \overline{L}$) is given by

$$\lambda n \overline{L} \leq \frac{\sqrt{8}}{\pi} \frac{1}{\Delta} W \sqrt{nA} - n\eta U(\epsilon^2) - nrU^{(b)}(\delta) \;\; \text{bit-meters/second} \tag{113}$$

Theorem 5 has interesting implications. The raw transport capacity of a wireless network scales as $\sqrt{n}$ while the overhead incurred by the routing overheads scales as $n$. Therefore if the number of nodes deployed in a network increases beyond a certain threshold, denoted by $n^\star$, then no useful information may be transported in the network and the whole capacity is used up by the geographic routing overheads. This leads to the following corollary.

*Corollary 3:* For geographic routing, the upper bound on the maximum number of nodes that may be deployed in a network while ensuring that it has non-zero residual transport capacity is given by

$$n^\star \leq \left( \frac{\frac{\sqrt{8}}{\pi} \frac{1}{\Delta} W \sqrt{A}}{\eta U(\epsilon^2) + rU^{(b)}(\delta)} \right)^2 \tag{114}$$

*Proof:* If the residual transport capacity is greater than zero then

$$\frac{\sqrt{8}}{\pi} \frac{1}{\Delta} W \sqrt{nA} - n\eta U(\epsilon^2) - nrU^{(b)}(\delta) \geq 0$$

which implies that

$$\sqrt{n} \left( \eta U(\epsilon^2) + rU^{(b)}(\delta) \right) - \frac{\sqrt{8}}{\pi} \frac{1}{\Delta} W \sqrt{A} \leq 0$$

Rearranging the above equation yields (114).                                                                                        ∎

Figures 6, 7, and **??** show the lower bound on the fraction of transport capacity that is used by the routing overheads. Figures 6 and 7 illustrate that for large network size and packet arrival rate, complete transport capacity of the network may be occupied by the routing overheads.

## VII. Discussions

In this section we briefly discuss some extensions of the rate-distortion formulation proposed in this paper. We first discuss the cases where our formulation and analysis yield trivial lower bound of zero. We then discuss how the formulation and analysis may be applied to other scenarios such as other location services, mobility models, etc. We then address some possible extensions to the model and analytical approach used in this paper.

### A. Trivial lower bound scenarios

The lower bound on the rate at which a node needs to update its location server, derived in Section IV, may be zero in certain cases. This lower bound is trivial and does not provide much useful information. In this subsection we provide a technique for obtaining more useful bounds for such cases. For clarity, we first consider the location update rate for one-dimensional networks and then explain how similar technique may be applied to two-dimensional networks. We also comment on cases where the results of Section V yield trivial lower bounds on minimum beacon overhead rate.

The lower bound on location overhead evaluated in IV-B is zero when $h(X_{i1}(T_1)) \leq \frac{1}{2} \log \left(2\pi e \epsilon^2\right)$. This occurs when the second moment of the change in position of a node during a packet inter-arrival period is less than the fidelity criterion $\epsilon^2$. This may be the result of low mobility (small $\sigma^2$) or average packet inter-arrival time (small $E[S]$). Although average change in position of a node during a packet inter-arrival time may be small, the position change accumulates over time. Thus instead of updating the location server between every packet inter-arrival duration, the nodes may need to update the location information once in a few packet inter-arrival durations. Thus we modify the problem formulation for minimum location update overhead to take this into account. The distortion criterion considered in Section IV is that the expected squared error in location information of a node must be less than $\epsilon^2$ at every time instant a packet destined to the node is generated in the network. We relax the distortion measure such that the squared error in location information is required to be less than $\epsilon^2$ for every $k^{th}$ generated for the node.

Let $f_S^k(t)$ denote the pdf of $T_k$, the time at which $k^{th}$ packet destined to a node is generated in the network. $T_k$ is simply the sum of $k$ packet inter-arrival durations, each of which is independently and identically distributed according to $f_S(t)$. Thus $f_S^k(t)$ is given by

$$f_S^k(t) = f_S^{k-1}(t) \star f_S^1(t) = \int_0^\infty f_S^{k-1}(t-\tau) f_S^1(\tau) d\tau \tag{115}$$

$$f_S^1(t) = f_S(t) \tag{116}$$

where $\star$ is the convolution operator. The pdf of $X_{i1}(T_k)$, denoted by $f_{X_1}^k(x)$ is given by

$$f_{X_1}^k(x) = \int_0^\infty \frac{1}{\sqrt{2\pi\sigma^2\tau}} e^{-\frac{x^2}{2\sigma^2\tau}} f_S^k(\tau) d\tau \tag{117}$$

and $H(X_{i1}(T_k))$ is given by

$$H\left(X_{i1}(T_k)\right) = -\int_{-\infty}^\infty f_{X_1}^k(x) \log \left(f_{X_1}^k(x)\right) dx \tag{118}$$

Let $k^\star$ be defined as

$$k^\star \triangleq \min_k k \in \left\{ k | H\left(X_{i1}(T_k)\right) > \frac{1}{2} \log \left(2\pi e \epsilon^2\right) \right\} \tag{119}$$

In other words, $k^\star$ is the minimum number of packet inter-arrival durations after which the second moment of the change in position of a node is greater than $\epsilon^2$. The value of $k^\star$ will depend on $\sigma^2$ and the distribution of packet inter-arrival duration. So if we consider the relaxed distortion measure where we are only concerned with the squared error in location information after every $k^\star$ packet generation instances, then following the same steps as in Subsection IV-B we get the following lower bounds

$$R\left(\epsilon^2\right) \geq \frac{1}{k^\star} \left( H\left(X_{i1}(T_{k^\star})\right) - \frac{1}{2} \log \left(2\pi e \epsilon^2\right) \right) \text{ bits/packet} \tag{120}$$

$$U\left(\epsilon^2\right) \geq \frac{1}{k^\star E[S]} \left( H\left(X_{i1}(T_{k^\star})\right) - \frac{1}{2} \log \left(2\pi e \epsilon^2\right) \right) \text{ bits/second} \tag{121}$$

Note that the right hand side of the above constraints is always positive and hence are more meaning for the cases where results of Section IV yield trivial bounds. In fact the bounds evaluated in Section IV are special cases of (120) and (121) corresponding to $k^\star = 1$.

Similarly for two-dimensional network we can find more meaningful bounds by considering the change in location during $k$ packet inter-arrival periods. An analogous $k^\star$ may be defined as

$$k^\star \triangleq \min_k k \in \left\{ k | H\left(X_{i1}(T_k)\right) + H\left(X_{i2}(T_k)\right) > \log\left(\pi e \epsilon^2\right) \right\} \tag{122}$$

For the relaxed distortion constraints we thus get the following bounds for the two-dimensional case

$$R\left(\epsilon^2\right) \geq \frac{1}{k^\star}\left(H\left(X_{i1}(T_{k^\star})\right) + H\left(X_{i2}(T_{k^\star})\right) - \log\left(\pi e \epsilon^2\right)\right) \text{ bits/packet} \tag{123}$$

$$U\left(\epsilon^2\right) \geq \frac{1}{k^\star E[S]}\left(H\left(X_{i1}(T_{k^\star})\right) + H\left(X_{i2}(T_{k^\star})\right) - \log\left(\pi e \epsilon^2\right)\right) \text{ bits/second} \tag{124}$$

The lower bound on the minimum beacon overhead, evaluated in Section V, is zero when $\mathcal{H}(p(l^\star)) \leq \mathcal{H}\left(\frac{\delta}{2}, 1 - \delta, \frac{\delta}{2}\right)$. This may happen if the $\sigma^2 E[\tau] >> 2r$. When $\sigma^2 E[\tau] >> 2r$, with probability greater than $1 - \delta$ a node will leave the neighborhood of its neighbors within a packet inter-arrival period. According to our formulation the node does not need to send a beacon as the neighbors may simply assume that it has left their neighborhoods. As a result the bound may be loose in such circumstances because it is inevitable that the nodes will need to exchange beacons in order to maintain consistent neighborhood information. The idea of relaxing the distortion constraint by considering accuracy of neighborhood information over several packet arrival intervals, as done for the location update case, does not work in this case. This is because as we increase the number of packet arrival intervals the probability that two nodes remain neighbors over that time duration decreases. However if the fidelity criterion is very strict ($\delta \approx 0$) then the bounds will be meaningful over a wide range of $\sigma$ and packet arrival rate distributions.

### B. Application of the formulation to other scenarios

In this section we comment on generality of the rate-distortion model and how it may be easily extended to other scenarios.

*1) Per-Session Location Discovery:* The network model used in this paper assumed that the location server of the destination is queried at the arrival of each new packet. Another implementation of geographic routing is possible where a source node queries the location server only on the arrival of a new session. During the session, the source and destination may piggy-back their location information along with the data and ACK packets respectively. Of course such a scheme would work well only if the session involves flow of traffic in reverse direction (destination to source) and the time elapsed between arrival of packets in the session is not large.

The overhead incurred in updating the location server for such a geographic routing protocol would be different from the results presented in Section IV. Also other than the overhead associated with updating the location server, another overhead associated with sending the location update information along with every ACK/data packet is also introduced.

The rate-distortion analysis of Section IV may be easily extended to analyze the overhead incurred by geographic routing with per-session location discovery. Let $S_1$ denote the time interval elapsed between arrival of two sessions destined for the same destination and let $S_2$ denote the inter-packet arrival time within a session. Let $f_{S_1}(t)$ and $f_{S_2}(t)$ denote the pdfs of $S_1$ and $S_2$ respectively. sing similar approach as Theorem 1, the overhead associated with updating the location server is greater than equal to

$$h(X_{i1}(T_1)) - \frac{1}{2}\log\left(2\pi e \epsilon^2\right) \text{ bits/session} \tag{125}$$

where

$$f_{X_1}(x) = \int_{\tau=0}^{\infty} \frac{1}{\sqrt{2\pi\sigma^2\tau}} e^{-\frac{x^2}{2\sigma^2\tau}} f_{S_1}(\tau) d\tau \tag{126}$$

Similarly the overhead incurred in sending the location information along with ACK/data packets is greater than equal to

$$2\left(h(f'_{X_1}(x)) - \frac{1}{2}\log\left(2\pi e \epsilon^2\right)\right) \text{ bits/packet} \tag{127}$$

where

$$f'_{X_1}(x) = \int_{\tau=0}^{\infty} \frac{1}{\sqrt{2\pi\sigma^2\tau}} e^{-\frac{x^2}{2\sigma^2\tau}} f_{S_2}(\tau) d\tau \tag{128}$$

The factor of 2 appears in the above expression because both the source and destination need to send their location information piggybacked with each data and ACK packet. The overall overhead incurred in maintaining location information is given by

$$U(\epsilon^2) \geq \frac{1}{E[S_1]} \left( h(f_{X_1}(x)) - \frac{1}{2}\log\left(2\pi e\epsilon^2\right) \right) + \frac{2}{E[S_2]} \left( h(f'_{X_1}(x)) - \frac{1}{2}\log\left(2\pi e\epsilon^2\right) \right) \text{bits/second} \tag{129}$$

The total transport capacity deficit is greater than equal to

$$\frac{n\eta}{E[S_1]} \left( h(f_{X_1}(x)) - \frac{1}{2}\log\left(2\pi e\epsilon^2\right) \right) + \frac{2n\eta'}{E[S_2]} \left( h(f'_{X_1}(x)) - \frac{1}{2}\log\left(2\pi e\epsilon^2\right) \right) + nrU^{(b)}(\delta) \text{bit} - \text{meters/second} \tag{130}$$

*2) Other Location Services:* The network model used in this paper accounts for a location service where the location of each destination is maintained by a single location server. Often location services use multiple location servers for maintaining location information of each destination. It should however be noted that the minimum rate at which a node must transmit its location information does not change and is given by (25). The multiple location servers affects the distance traveled by this location information generated by a node.

Consider a location service that assigns $k$ location servers for each node. Each of the $k$ location servers is updated by a node using $k$ independent messages. Let $\eta_i$ be the average distance between a node and its $i^{th}$ location server. The transport capacity deficit caused by a such a scheme would be greater than equal to

$$\left( \sum_{i=1}^{k} \eta_i \right) nU\left(\epsilon^2\right) + nrU^{(b)}(\delta) \text{ bit} - \text{meters/second} \tag{131}$$

In several location service schemes independent update messages are not issued. Instead the update messages may be routed along a multicast tree so that all the location servers are updated. In such a case, the distance traveled by the location update messages, $\eta$, would be equal to the length of the multicast tree. For example, consider XYLS (also known as column-row location service) [28]. In XYLS, each node maintains its location at every node that lies within the column containing the node. This is done by sending an update message in the north-south direction. Every node that overhears the messages updates the location information. Thus in a square field of area $A$ meter$^2$, a location update bit would travel $\sqrt{A}$ meters, i.e., $\eta = \sqrt{A}$. Thus the transport capacity deficit caused is greater than equal to

$$\sqrt{A}nU\left(\epsilon^2\right) + nrU^{(b)}(\delta) \text{ bit} - \text{meters/second} \tag{132}$$

*3) Hierarchical Location Services and Distance Effects:* In hierarchical location service schemes, there exists a hierarchy of location servers [29], [30]. The level of the server in the hierarchy depends on the distance from the node it serves. The location servers closest to a destination, and lowest in the hierarchy, have the most accurate information about the destination. As we move up the hierarchy, the location information becomes less accurate. That is farther a location server is from the destination it serves, less accurate would be the location information maintained by it. This scheme works well due to the *distance effect* - greater the distance from a node, slower it appears to move.

Let $\epsilon_i$ be the expected error in location information available with the location servers at level $i$. Let $\eta_i$ be the average distance traveled by bits in order to update the servers at level $i$. Thus the transport capacity deficit for such a hierarchical location service scheme is greater than equal to

$$n \sum_{i=1}^{k} U\left(\epsilon_i^2\right) \eta_i + nrU^{(b)}(\delta) \text{ bit} - \text{meters/second} \tag{133}$$

*4) Other Mobility Models:* In this paper we use Brownian motion to model node mobility. Although Brownian motion may not be the most realistic mobility model, the existence of closed form expression for node location pdf makes analysis tractable. Also Brownian motion allows us to change the degree of uncertainty in node position by changing a single parameter $\sigma$.

Many other mobility models for ad hoc networks have been proposed [31], [32], some of which claim to be a better reflection of reality than the other. Our analysis may be easily extended to any other mobility model, as long as the expression for steady state probability density function of the node locations is know. The pdf of node location may be plugged into (27) in order to obtain $f_{X_1}(x)$. However it is a non-trivial exercise to evaluate the closed form expression of the node location for mobility models like random way-point model, Gauss-Markov model, etc.

### C. Extensions to the Model and Analysis

In this subsection we discuss some of the possible extensions to the model and analysis that may be incorporated in the future.

*1) Incorporating the Caching of Location Information:* The network model does not take into account the caching of location information at the source and intermediate nodes. For example, the source node may cache the location information of the destination received from the location server and may add the cached information to the headers of future packets. The cached information may be used until it expires after some fixed time. Also intermediate nodes with fresher information regarding the location of the destination may update the packet header. Such a caching scheme has not been incorporated in our model. Caching may save the overhead associated with periodically querying the location server, although the location update and beacon overheads may remain unchanged.

*2) Exploring Closed Form and Tighter Bounds:* Although the network model used in this paper is very simple, still it is not possible to obtain closed form analytical expressions for many quantities. For example, a closed form expression of minimum overhead incurred by location update packets cannot be derived if the packets arrival process is Poisson. Similarly for the minimum beacon overhead case, all the results are expressed in terms of $\mathcal{H}\left(p\left(l^{\star}\right)\right)$. However numerical methods may be easily applied to evaluate the values of the expressions for given parameters. Finding tight closed form bounds for the expressions would also be the focus of future work.

### VIII. CONCLUSION AND FUTURE WORK

In this paper we studied the protocol overhead incurred by geographic routing in order to route packets with a given level of reliability. The protocol overhead is categorized into *location update overhead* and *beacon overhead*. For both kind of overheads, the problem of finding minimum overhead incurred is formulated as rate-distortion problem. For location updates we evaluate a lower bound on the minimum rate at which a node must transmit its location information to the location server so that the expected error in location information used for routing is less than a given value. For the beacon updates, we evaluate a lower bound on the minimum rate at which a node must transmit beacon packets so that the probability that its neighbors maintain a correct neighborhood information is greater than a given value. We first evaluate the bounds for one-dimensional networks and then extend the results to two dimensional networks. We also characterize the deficit in capacity caused by the routing overheads.

Developing an information theoretic framework for evaluating the minimum protocol overhead incurred by several classes of routing protocols, such as proactive, reactive, and hierarchical routing protocols, is the ultimate goal of the framework proposed here. Such a universal analytical framework for characterizing routing overheads for all routing paradigms would be useful in determining which protocol is suitable for a given scenario. Development of efficient routing protocols for mobile ad hoc networks, using the information theoretic results as a guideline, would also be a part of future research.

### ACKNOWLEDGMENT

This work was funded in part by the National Science Foundation under grants CNS-322956 and CNS-546402.